\documentclass[a4paper,11pt]{article}
\usepackage[]{amsmath,amssymb}
\usepackage{color}
\usepackage{hyperref}
\usepackage{graphicx}
\usepackage{siunitx}
\usepackage[nottoc]{tocbibind}

\usepackage{subcaption}
\usepackage{mwe}
\usepackage[textwidth=17cm,textheight=23cm]{geometry}

\usepackage{authblk}

\begin{document}

\title{Vibrational coherent control of localized d-d electronic excitation}

\author[1,2,$^{\star}$]{Alexandre Marciniak}
\author[1,3,$^{\star}$]{Stefano Marcantoni}
\author[1,2]{Francesca Giusti}
\author[1,2]{Filippo Glerean}
\author[1,2]{Giorgia Sparapassi}
\author[5]{Tobia Nova}
\author[5]{Andrea Cartella}
\author[5]{Simone Latini}
\author[1]{Francesco Valiera}
\author[5]{Angel Rubio}
\author[4]{Jeroen van den Brink}
\author[1,3]{Fabio Benatti}
\author[1,2,6,$^{\dagger}$]{Daniele Fausti}

\affil[1]{Department of Physics, University of Trieste, Via A. Valerio 2, 34127 Trieste, Italy}
\affil[2]{Elettra-Sincrotrone Trieste S.C.p.A. Strada Statale 14 - km 163.5 in AREA Science Park 34149 Basovizza, Trieste, Italy}
\affil[3]{National Institute for Nuclear Physics (INFN), Trieste Section, I-34151, Trieste, Italy}
\affil[4]{Leibniz Institute for Solid State and Materials Research, 01069 Dresden, Germany}
\affil[5]{Max Planck Institute for the Structure and Dynamics of Matter, Hamburg, Germany}
\affil[6]{Department of Chemistry, Princeton University, Princeton, New Jersey 08544, United States}
\affil[$^{\star}$]{Those authors have contributed equally to this work}
\affil[$^{\dagger}$]{Corresponding author : \href{mailto:daniele.fausti@elettra.eu}{daniele.fausti@elettra.eu}}

\date{  }
\maketitle

%\begin{bibunit}[unsrt]
\baselineskip=18pt
\section*{Abstract}
\addcontentsline{toc}{section}{Abstract}
\textbf{Addressing the role of quantum coherence in the interplay between the different matter constituents (electrons, phonons and spin) is a critical step towards understanding transition metal oxides and design complex materials with new functionalities. Here we use coherent vibrational control of onsite d-d electronic transitions in a model edge-sharing insulating transition metal oxide (CuGeO$_3$) to single-out the effects of vibrational coherence in electron-phonon coupling. By comparing time domain experiments based on high and low frequency ultrashort pumps with a fully quantum description of phonon assisted absorption, we could distinguish the processes associated to incoherent thermal lattice fluctuations from those driven by the coherent motion of the atoms. In particular, while thermal fluctuation of the phonon bath uniformly increases the electronic absorption, the resonant excitation of phonon modes results also in light-induced transparency which is coherently controlled by the vibrational motion.}\\ 

The non-trivial interplay between high energy electronic transitions, both onsite and charge-transfer, with low energy lattice or magnetic excitations gives rise to the rich phase diagrams of transition metal oxides \cite{Dagotto2005,Zaanen1985}. Unveiling the details of this interplay is the key to achieve a better and more reliable description and design of material properties. We stress that the incoherent energy exchange rate between different degrees of freedom at play may not provide a complete physical picture of the coherent interaction between the electrons and the phonons (or magnons). The role coherences of low energy excitation play in determining macroscopic material properties is often elusive and addressed indirectly through a population dynamics of high energy excitations interacting with low energy modes. \\

In time domain studies, electron-phonon coupling is often addressed indirectly by inferring the strength of the coupling from the relaxation dynamics following a sudden photo-injection of electronic energy through high energy photon pulses \cite{DalConte2012}. In this setting vibrational and magnetic coherences can be studied only in cases subject mainly to two strong limitations: the sudden photoexcitation should be able to trigger a coherent vibrational (or magnetic) response and the interaction between the probe and the material ``prepared" in a coherent vibrational state should map such coherence into photonic observables. The resonant excitation of vibrational modes (commonly dubbed ``phonon pump") partially lifts these limitations as the creation of coherent lattice response is driven directly by the e.m. field pulse and not mediated by electron relaxation \cite{Forst2011,Mankowsky2014,Mankowsky2016}. Contrary to what happens in standard high photon energy pump and probe measurements, mid-IR excitation drives a large-amplitude low-frequency vibrational response through the resonant excitation of phonon-polariton modes. Under this condition, a coherent vibrational excitation is prepared in the electronic ground state of the material, providing the means to dynamically control the atomic position in matter \cite{Cartella2017}.\\ 

Here we use resonant vibrational excitation to coherently control the crystal field surrounding the Cu ions in a model compound for edge-sharing cuprate. The rationale of our work is the following. The mid-IR excitation resonantly excites a large amplitude motion of the ions mainly along an infrared active mode. The anharmonic coupling of the excited IR-phonon to other vibrational modes results in a coherent contraction and expansion of the Cu-O bonds, within the octahedra, which coherently control the absorption in the visible region due to onsite optical transitions between crystal field levels. In detail, the resonant excitation of IR active phonon modes results in a coherent vibrational motion of the apical oxygen which dynamically controls the energy and oscillator strength of orbital transition between different crystal levels on Cu$^{2+}$ ions. The details of the DFT estimation of anharmonic coupling are described in the Supplementary Information (SI, section \ref{section_DFT_calc}) and the effective force field acting on the apical oxygen is depicted in Fig. \ref{force_field} of SI. \\

The coherent vibrational control of the electronic transition is evidenced by the striking contrast between results of time domain experiments based on high photon energy pumps \cite{Giannetti2016} and mid-IR excitation. While high photon energy excitation results in thermal disorder that uniformly increases the absorption of crystal field levels \cite{Giannetti2009,Yuasa2008}, our experiments based on mid-IR pumps reveals a transient response characterized by regions of induced transparency which can only be rationalized if the electronic transitions are dynamically controlled by vibrational coherence in the electronic ground state. In order to disentangle the contributions to crystal field absorption which result from coherent and thermal motion of the ions, we developed a fully quantum description of dynamical phonon-mediated crystal field excitations. We use the temperature dependent equilibrium absorption to benchmark the role of thermal fluctuation in the absorption process and extract a quantitative description of coherent vs incoherent vibrational responses. We stress that our methodology allows for the first time to distinguish the contribution to the absorption of crystal field levels which are associated to the coherent motion of the ions from the one driven by their incoherent thermal fluctuations.\\

\begin{figure}[h!]
\centering
\includegraphics[scale=0.56]{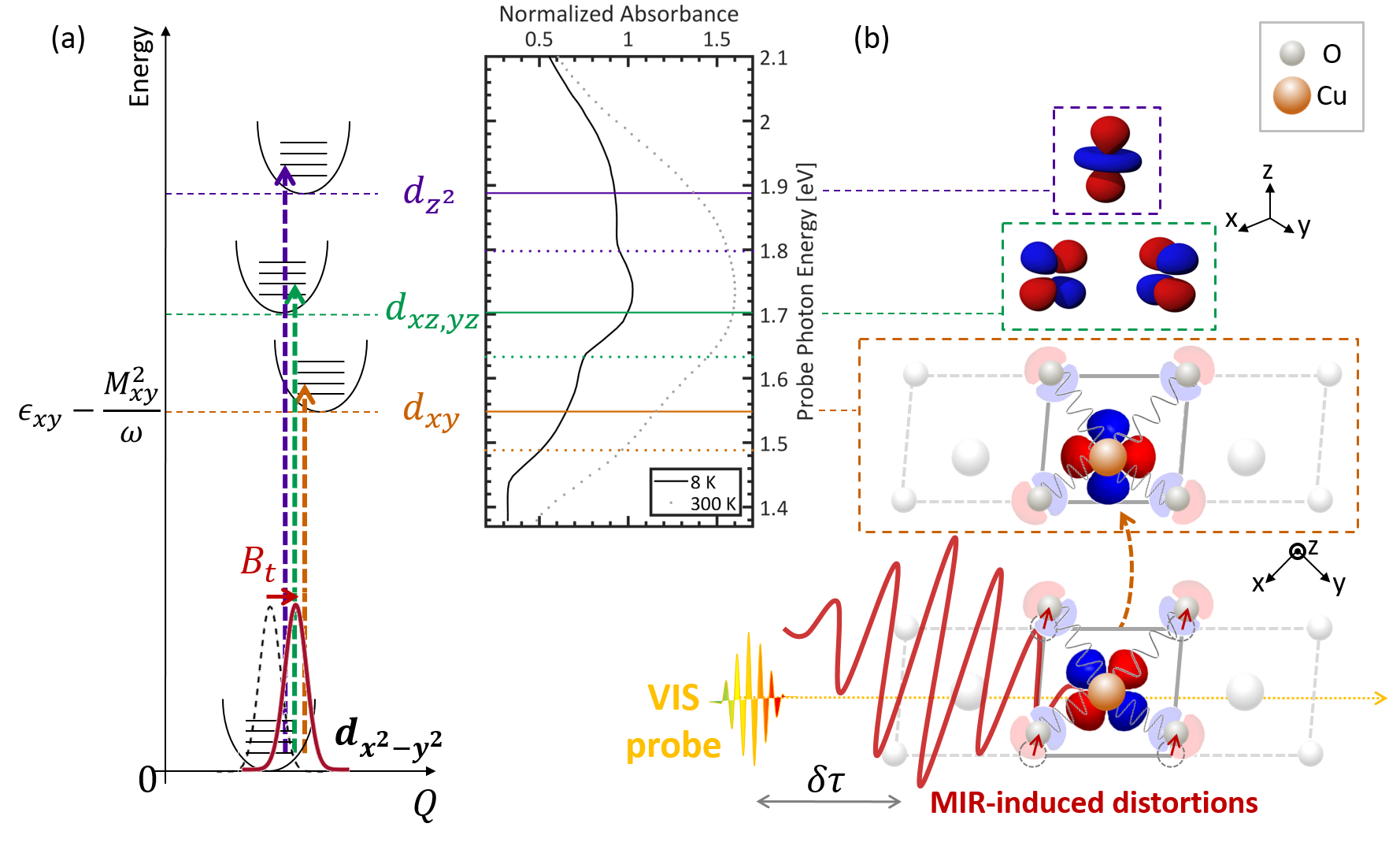}
\caption{\textbf{Coherent vibrational control of onsite d-d crystal field transitions between different Cu orbital states.} (a) The coupling of onsite transition between different orbital states of Cu$^{2+}$ to low energy phonon in the ground state is indicated by the temperature dependent static absorbance (insert) which consists of three main features associated to electronic transitions from the electronic ground state orbital d$_{x^2-y^2}$ to the excited states d$_{xy}$ (1.55 eV), d$_{xz}$, d$_{yz}$ (1.7 eV) and d$_{z^2}$ (1.88 eV) whose oscillator strength increases with temperature. (b) In our experiment, a mid-IR pump pulse excites vibrational modes displacing, through non-linear phononic coupling, the atomic position along the normal coordinate axis (B$_t$). This distortion couples to the electronic wavefunction and induces oscillator strength variations of the Cu on-site d-d orbital transitions which are measured by a delayed visible pulse whose central energy is tuned across the d-d transition energies.}\label{Scheme_of_principle}
\end{figure}

The measurements are performed in a model system for edge-sharing cuprate, insulating CuGeO$_3$. This sample is ideal for two main reasons: (i) the phonons are long lived allowing therefore for a selective excitation of vibrational modes and (ii) the three d-d electronic transitions at high energy (around 1.7 eV) are isolated from other spectral features such as the charge transfer edge \cite{Damascelli2000} (see Supplementary - \nameref{sample_properties}). The relevant structural unit for our discussion is a CuO$_6$ complex, with the copper atom surrounded by six oxygens at the vertices of a distorted octahedron. The three observed optical absorptions (insert in Figure \ref{Scheme_of_principle}.a) are due to onsite electronic transitions between different d-orbitals of copper whose degeneracy is removed by the broken octahedral symmetry. In particular the three transitions observed are transition from the ground state (d$_{x^2-y^2}$) to in plane d$_{xy}$ orbital and out of plane d$_{xz}$, d$_{yz}$, d$_{z^2}$ (see Supplementary).\\ 

Importantly, these optical transitions are phonon-assisted. The onsite optical transition between orbitals of d symmetries should be forbidden in centrosymmetric crystals due to dipole selection rules ($\Delta \ell$ = 0), nevertheless, they are visible in absorption spectroscopy even at very low temperatures ($<$ 10K) \cite{Bassi1996,ONeal2017}. This can be understood by considering that the optical onsite d-d transitions are accompanied by the creation (and annihilation) of phonons which break the symmetry of the orbital transition thereby removing the optical selection rule \cite{Bassi1996,ONeal2017,Monney2013}. The involvement of a ground state phonon mode in the optical absorption process is confirmed by the observed strong increase of the oscillator strength with temperature \cite{Bassi1996}.\\

We describe the transient coherent response in terms of an effective model for the phonon-mediated onsite absorption where we consider the d-d transition as a two-level system interacting with a vibrational mode which is represented by a quantum harmonic oscillator \cite{Mahan2000}. We choose this minimal modelling, which contains only one electronic transition and one phonon mode, because it is simple enough and we can treat the electric field pulse with a full quantum formalism. We will show that this simple model contains the relevant features to grasp the basic physical mechanism of vibrational coherent and incoherent (thermal) control of d-d absorption. The material effective Hamiltonian is therefore: 

\begin{equation}
H = \omega b^{\dagger} b+\epsilon d^{\dagger} d + M d^{\dagger} d(b+b^{\dagger}) ~,
\end{equation}

\noindent where $b$ (and $b^{\dagger}$) are bosonic operators for the phonon mode (of energy $\omega$, $\hbar=1$), $d$ (and $d^{\dagger}$) are fermionic operators describing the onsite d-d electronic transition (of energy $\epsilon$) and the last term describes the interaction between the d-d electronic transition and the phonon displacement (with coupling strength $M$). In order to describe the coherent and incoherent vibrational dressing of onsite crystal field transition we can consider the interaction between an external electric field and the onsite d-d transition coupled to low energy phonons using the following interaction Hamiltonian:

\begin{equation}\label{eq_Hint}
H_{int} = \mu_0 (b+b^{\dagger})(d+d^{\dagger}) \Sigma_k(a_k+a_k^{\dagger}) ~,
\end{equation}

\noindent where the bosonic operators $a_k$ describe the electric field operators at frequency $\nu_k$ interacting with the sample. The electric field operator can be used to describe an incoherent field as in linear response or the spectral components of a probe pulse in time domain experiments. With this model we can retrieve the total absorption of phonon assisted d-d transitions at equilibrium which depends on the temperature of the system as revealed by equilibrium absorption measurements of the CuGeO$_3$ as a function of the temperature \cite{Bassi1996,ONeal2017} (see SI for the analytical derivation of the temperature dependence  eq. \ref{abso1}, \cite{Ballhausen1962}). Importantly, our formalism describes also the non-zero absorption at T = 0 K and validates the phonon-assisted character of the three d-d transitions observed in optical spectroscopy (Fig.1) which, in details, are the in plane transition d$_{x^2-y^2} \rightarrow$ d$_{xy}$  and the out of plane transitions d$_{x^2-y^2} \rightarrow$ d$_{xz}$, d$_{yz}$  and d$_{x^2-y^2} \rightarrow$ d$_{z^2}$.\\

A scheme of our experiment is depicted in Figure.\ref{Scheme_of_principle}. In short, a mid-IR pulse excites a phonon mode that drives a displacement of the ionic structure. This in turn modifies the interaction energy between the lattice and the orbital degrees of freedom that are then investigated by a visible probe pulse. Since d-d electronic transitions in CuGeO$_3$ have different symmetry, the expected ultrafast variations of their oscillator strength and central energy are related to the orbital coupling to the structural changes induced by the phonon pump.\\

In order to reveal these effects, we have developed an optical setup that produces mid-IR pump pulses (tunable from 5 µm to 18 µm) coupled with visible probe pulses ($\approx$ 30 fs and tunable from 650 nm, 1.91 eV, to 950 nm, 1.30 eV). The detection is made by a low-noise balanced photodiode coupled to a fast digitizer system which measures the transient intensity of the probe transmitted through the mid-IR-excited CuGeO$_3$ single crystal fixed in a cryostat kept at 8 K. All polarizations were oriented along the c-axis of CuGeO$_3$ which is the direction of the CuO$_6$ octahedron chain (see Supplementary - \nameref{experimental_method}). 

The pump wavelength dependence of the transient transmissivity is reported in Figure 2a for probe energy matching the center of the d-d band ($E_{probe}\approx$ 1.75  eV). Similarly to previous studies \cite{Giannetti2009,Yuasa2008}, we observe that the change in transmission is negative, indicating that the d-d transition oscillator strength increases following the photoexcitation. Two types of dynamics can be distinguished: a broadband fast one and a slow one, whose rising time is about several ps, and is visible only for specific pump wavelengths (around $\lambda_{pump} \approx$ 7.7 $\mu$m and $\lambda_{pump} \approx$ 11 $\mu$m). The slow picosecond response can be understood as a thermal response of the material, which is maximized at pump energies where linear dissipative absorption can be observed, i.e. the slow response is maximized at pump wavelength where linear absorption is maximum. The fast and intense response is instead resonant to the \textit{reststrahlen} band associated to phonon modes visible in the linear optical conductivity (see Supplementary - \nameref{pump_wavelength_dependency}) and associated to the non-linear coherent phonon dynamics which we will discuss in the following.\\

The fast coherent response has its maximum amplitude around $\lambda_{pump} \approx$ 9 $\mu$m, which is out of resonance from any linear dissipative absorption of the phonon modes \cite{Damascelli2000,Popovic1995}. It resonates at the wavelength where the pump can drive the largest phononic inductive response. It is important to note that the resonant frequency is not exactly at the phonon reststrahlen band but volume and propagation effects play a role in determining the optimal pumping wavelength \cite{Cartella2017} (see Supplementary – \nameref{pump_wavelength_dependency}). In this condition, the resonant excitation of vibrational modes drives a large amplitude displacement of the ions which in turns triggers, through non-linear phononic coupling, a strong displacement of the ions along different phonon modes which is typically described in terms of non-linear phononic response with a minimal Hamiltonian for phononic coupling (for a description of non-linear phononic coupling of type $Q_{IR}^2 Q_{Raman}$, see Supplementary - \nameref{anharmonic_coupling}). \\

\begin{figure}[h!]
\centering
\includegraphics[scale=0.3]{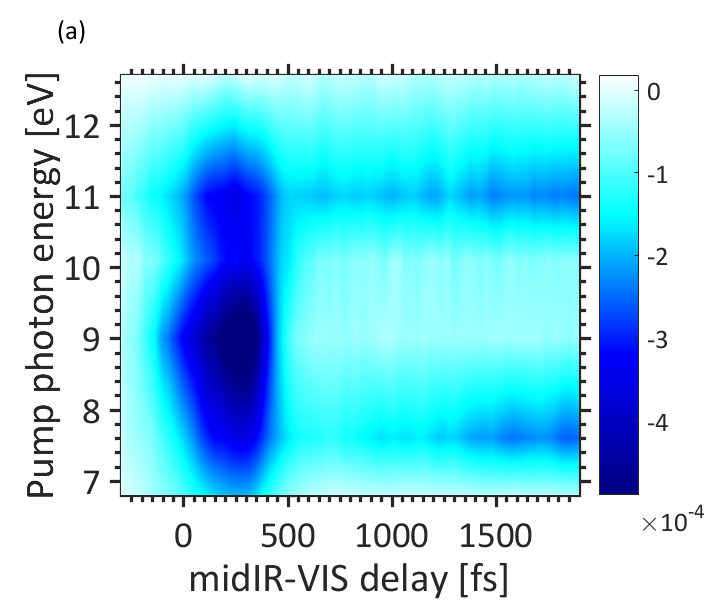}\includegraphics[scale=0.3]{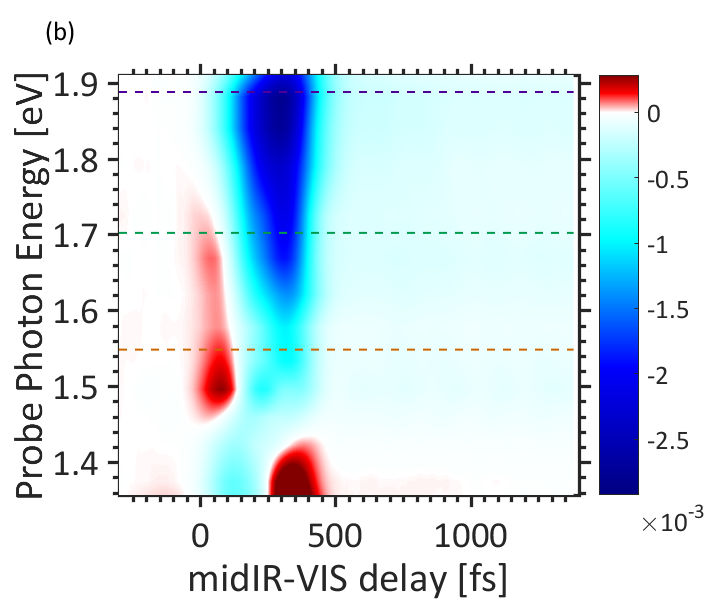}\includegraphics[scale=0.3]{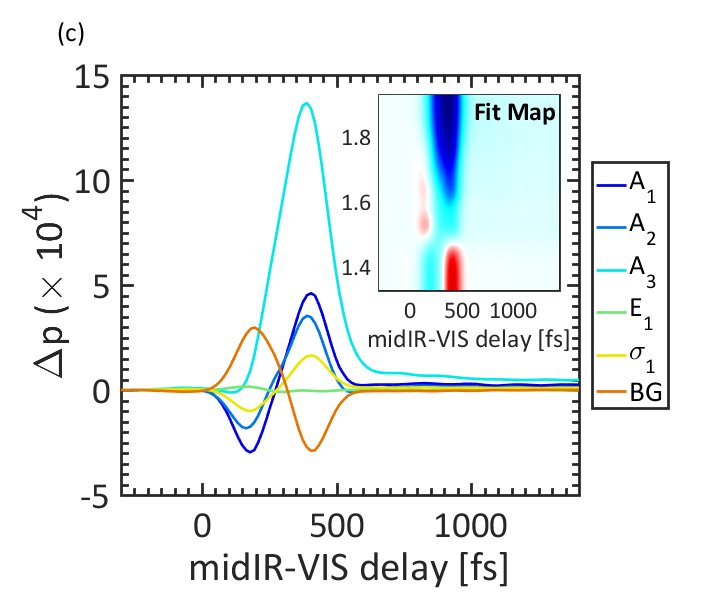}
\caption{\textbf{Experimental evidence of coherent and incoherent phonon dressings of d-d crystal field transitions.} (a) Pump wavelength dependency of the transmissivity for $E_{probe} \approx$ 1.7 eV. The color scale is slightly saturated in order to better observe the long timescale dynamics associated to a dissipative thermal response (see text). (b) Transmissivity as a function of the probe photon energy and of the pump-probe delay, measured for resonant pump wavelength ($\lambda_{pump} = 9$ $\mu$m). A positive signal is measured instantaneously after the excitation, around the two first transitions, and at $\tau$ = 400 fs in the lowest energy area. Dashed lines highlight the central energies of the three crystal field levels. (c) Transient Gaussian fitting parameters of the three d-d absorption peaks: amplitude ($A_i$), central energy ($E_i$) and bandwidth ($\sigma_i$) and background ($BG$) (fit results in the insert).}\label{Experimental_results}
\end{figure}

In order to disentangle the coherent from incoherent phononic dressing of the d-d electronic transitions of CuGeO$_3$, we have measured the probe transient transmissivity as a function of probe photon energy, by keeping $\lambda_{pump}$ centered at 9 $\mu$m and the sample at a base temperature T = 8 K (see Supplementary – \nameref{thermal_effects}, for measurements at 300 K). The measured response, plotted in Figure \ref{Experimental_results}.b, reveals for short timescales a very different transient response for different probe photon energy. \\

We will focus on the 0 to 500 fs range (for a discussion of the coherent phonon response observed at longer times, see Supplementary – \nameref{longtime}). On short timescale, we can observe three probe energy ranges giving rise to different time domain responses. For a probe energy below 1.45 eV, the transient transmissivity starts negative and switches to positive values around $\tau$ = 300 fs. On the contrary for probes at 1.45 eV to 1.75 eV, the transient transmissivity is at first positive and subsequently drops to negative values. Moreover, in that area, two energy substructures are visible around 1.5 and 1.7 eV. For a probe energy greater than 1.75 eV, the transient transmissivity is fully negative and it shows an uncommon temporal structure where the decreasing slope is longer (about 350 fs) than the following rising edge (about 150 fs). The photon energy dependence of the probe response is representative of the position of the three d-d transitions which central energies are at: 1.5 eV for d$_{x^2-y^2} \rightarrow$ d$_{xy}$, 1.7 eV for d$_{x^2-y^2} \rightarrow$ d$_{xz}$, d$_{yz}$ and 1.9 eV for d$_{x^2-y^2} \rightarrow$ d$_{z^2}$ \cite{Bassi1996,Huang2011}.\\
 
The low energy region (probe energy around 1.45 eV) displays a transient transparency which can be quantified via a differential fit based on the linear absorbance, which takes into account time-dependent variations of the d-d transition oscillator strength and central wavelength (Fig. \ref{Experimental_results}.c, see Supplementary Figure \ref{Fitting_procedure} for the fitting procedure). The best-fit is obtained by letting free the central energy and the bandwidth of the first transition (d$_{x^2-y^2} \rightarrow$ d$_{xy}$) and the background. Interestingly, this procedure suggests that (i) the background amplitude seems to rule the transient transparency below 1.45 eV and (ii) the central energy and bandwidth of the d$_{x^2-y^2} \rightarrow$ d$_{xy}$ account for the dynamically changed response at higher energies.

In order to describe coherent and incoherent contribution to the time domain response of onsite d-d transition we consider the initial vibrational state (before the probe arrives) as a displaced thermal state. In this effective language, resonant excitation of the vibrational IR mode that is non-linearly coupled to the relevant octahedral vibration controlling d-d transitions is caused by a displacement operator, $D = e^{B_t b^{\dagger} - B_t b}$, acting on an initial thermal vibrational state. The total absorption (computation detailed in Supplementary – \nameref{theory}) reads:

\begin{equation}\label{eq_full_quantum_model}
\Gamma_{tot}^{displaced} = \mu_0^2 \left[ 4 B_t^2 + \coth\left( \frac{\beta \omega}{2} \right) \right] ~,
\end{equation}

\noindent where $B_t$ corresponds to the time dependent displacement along the phonon normal mode of energy $\omega$, $\mu_0$ is the oscillator strength and $\beta$ is the inverse temperature ($1/k_B T$). Note that this result is consistent with the empirical description given in ref. \cite{ONeal2017} for the temperature-dependent d-d band absorption measured.\\ 

\begin{figure}[h!]
\centering
\includegraphics[scale=0.4]{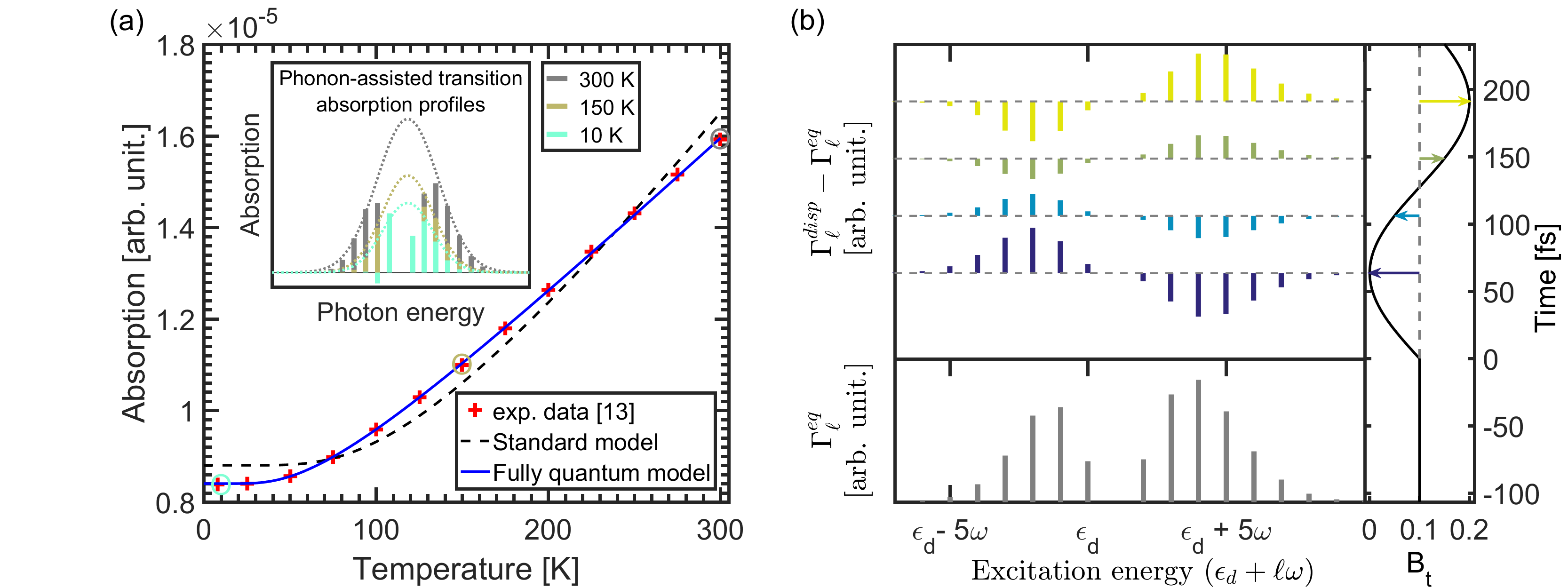}
\caption{\textbf{Phonon mediated crystal field absorption.} (a) Temperature dependence of the total optical absorption associated to the onsite d-d transitions. The absorption data (red cross) are taken from \cite{ONeal2017}. The ground state distortion ($B_{t,eq}$) revealed by our full quantum model (eq.\ref{eq_full_quantum_model}) gives a temperature dependence of the oscillator strength (blue line) in good agreement with the experimental data (the dash line is a standard model including solely the coupling with the phonon proposed and no feedback on the atomic position). The insert shows the equilibrium absorption profile for three different temperatures (note that the Gaussian simulates the broadening of the solid-state phase). (b) The variation of the absorption spectral line distribution ($\Gamma_{\ell}^{disp} - \Gamma_{\ell}^{eq}$ for four distinct normal displacements along the coherent vibrational motion at different times indicate that the coherent vibrational motion can coherently control on site d-d electronic transition in the visible.}\label{Theoretical_results}
\end{figure}

As reported in Fig. \ref{Theoretical_results}.a, our model accounts for all contributions to equilibrium d-d absorption including both thermal and quantum fluctuations of the atomic positions and identifies the frequency of the boson mediating the coupling to be $\omega$ = 131 cm$^{-1}$, with a corresponding equilibrium displacement $B_{t,eq}$ = 0.62  [arb. unit.]. We point out that the extracted values are taking into account the overall d-d band absorption, which is not representative of a single transition and limit our description to one phonon mode leading possibly to an overestimated $B_{t,eq}$ (see Supplementary Figure \ref{T_B_dependencies_large_scale}). In order to tune our model parameters, we can compare the computed equilibrium absorption as a function of the temperature (insert of Figure \ref{Theoretical_results}.a). Note that the discrete spectral lines are expected in the case of a molecular system, while, in the solid state context, the lineshape will be smeared out by the presence of a non-flat band structure, inhomogeneous and homogeneous broadening. Thus, we show that an increase of the temperature mostly increases all the discrete transitions which leads to a global absorption increase. \\

Interestingly, the time dependence of the atomic position ($B_t$) produced by mid-IR excitation and phonon non-linear coupling leads to a different behavior for the phonon assisted absorption probabilities $\Gamma_{\ell}$ (computation detailed in Supplementary – \nameref{theory}), where the parameter $\ell$ being the electronic transition accompanied by the production of $\ell$ phonons. In Figure \ref{Theoretical_results}.b, we display the variations of the transition probability with respect to the equilibrium distribution ($\Gamma_{\ell}^{disp} - \Gamma_{\ell}^{eq}$) in the case of four distinct displacements away from the equilibrium position. These profiles represent the changes in absorption induced by coherent vibrational motion at different time and show that the absorption central frequency can be shifted up or down in energy and result in regions of vibrationally controlled transparency, as observed in experiments. A quantitative agreement of the central frequency and energy bandwidth dynamics of the absorption line profile can be extracted from first and second momentum of the distribution \cite{vandenBrink2001} and results in elastic displacement of $\Delta B_t = 10^{-3} \cong 10^{-4}$ \AA ~ and changes in central frequency of the electronic absorption  $\Delta \langle E \rangle \approx 60$ $\mu$eV and bandwidth $\Delta \sigma \approx 18$ $\mu$eV (see Supplementary for detailed comparison between theoretical calculation and experimental results).\\

The paradigm of most of the pump and probe studies is to photo-excite at high frequency and to consider an energy flow from electrons to phonons. The example of CuGeO$_3$ is emblematic in this respect. In standard photo-doping experiments the excess of electronic energy injected by the pump is quickly redistributed (few tens of fs) toward phonon modes and it induces a global heating of the system. This leads to an increased disorder which results in strengthening the phonon assisted crystal field absorption (eq.\ref{eq_Hint} and Fig. \ref{Theoretical_results}.a) and reference \cite{Giannetti2009,Yuasa2008}.\\

The coupling processes between electrons and phonons are often described by an effective-coupling between electrons and a bath of phonons which is kept in thermal equilibrium. In this condition the coupling between the material and the electromagnetic field is described by a Fermi golden rule where the absorption cross section is obtained by a dipole operator connecting two eigenstates of the system with definite electronic and phononic excitations. Importantly, the non-equilibrium evolution is normally described by a simple extension of this approach where the temperature of phonons and electrons is allowed to change independently in time. This accounts for a large part of our experiments because any injection of energy should increase the phononic temperature and its disorder enhancing the dipole-forbidden d-d transition probabilities. \\

Nevertheless, we stress that any effective-temperature approach will fail in describing the coherent response observed when pumping low-energy degrees of freedom with the mid-IR pulse and probing electronic d-d transitions in the visible. The coherent control of transmissivity at short time scales ($<$ 500 fs) revealed a complex probe energy dependency (see Fig. \ref{Experimental_results}.b). In particular, the d$_{x^2-y^2} \rightarrow$ d$_{z^2}$ transition is continuously made more absorptive by the mid-IR excitation and quickly recovers the equilibrium value when the exciting field is gone. More interestingly both transitions d$_{x^2-y^2} \rightarrow$ d$_{xy}$, and d$_{x^2-y^2} \rightarrow$ d$_{xz}$, d$_{yz}$ show a transient transparency at short times that cannot be described by an increase of the phonon temperature and indicate a coherent vibrational control of the electronic transition probabilities. \\

The optical absorption dynamic is the result of coherent lattice distortion along different phonon modes of the system. Our model explains this fact as the result of a displacement of the ions in the electronic ground state along a direction coupled to the electronic transition, which could not be described by standard multi-temperature models but requires a full treatment of the coherences of the low energy degrees of freedom.\\

In summary, we have demonstrated that vibrational pumping can be used to coherently control optical transitions of electronic origin. The mid-IR excitation of IR active phonon modes, together with a strong lattice anharmonicity, can be used to dynamically control the position (and momentum) of the atoms that in turn modifies the crystal field electronic transition in a model system for transition metal oxides. The experimental evidence of light induced transparency controlled by the coherent vibrational motion, supported by a simple theoretical model, provides the means to measure electron phonon coupling in complex materials with phase sensitivity with respect to the vibrational motion, i.e. beyond the population driven incoherent coupling description. In the context of superconductivity, our approach, which could be extended to more complex Hamiltonian interactions, may provide a guideline to experimentally address the gap between BCS, with non-local and instantaneous interactions, and Eliashberg approaches which are local in space and retarded in time. The possibility of driving coherent vibrational excitations and to control local electronic degrees of freedom may provide the means to address the coherent vs. incoherent contributions to the interactions between electrons and phonons (or spins) and to address directly the delay in the development of the overscreening of Coulomb repulsion at the core of most of the quantum coherent phases observed in transition metal oxides. 

\section*{Aknowledgments}
\addcontentsline{toc}{section}{Aknowledgments}
We gratefully thank Alexandre Revcolevschi for providing the CuGeO$_3$ sample and checking the manuscript. Moreover, we thank Andrea Cavalleri for his feedbacks on the manuscript. This work was supported by the ERC-grants INCEPT n°67748. Moreover, this work was supported by the European Research Council (ERC-2015-AdG694097), the Cluster of Excellence AIM and SFB925.

\begin{footnotesize}
\bibliographystyle{unsrt}
\bibliography{Merged_Article_SOM_Marciniak2020_Vibrational_Coherent_Control}
%\bibliography{Biblio_vibration_control_dd_CuGeO3}
\end{footnotesize}
%\end{bibunit}

%\begin{bibunit}[unsrt]
\newpage
\clearpage
\begin{center}
\begin{Huge}
Supplementary Information
\end{Huge}
\end{center}

\setcounter{section}{0}
\setcounter{equation}{0}
\setcounter{figure}{0}
\addcontentsline{toc}{section}{Supplementary Information}
\section{Experimental details}
\subsection{Sample properties}
\label{sample_properties}
\begin{figure}[h!]
\centering
\includegraphics[scale=0.4]{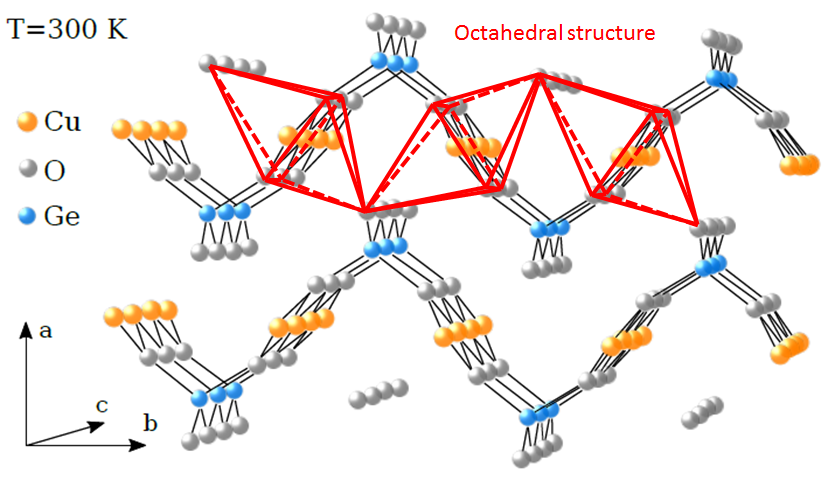}
\caption{Crystalline structure of CuGeO$_3$. The octahedral structure is underlined and it highlights the place of the Cu on-site d-d transitions. Figure adapted from \cite{Damascelli2000}.}\label{CuGeO$_3$_structure}
\end{figure}

Copper Gemanate (CuGeO$_3$, see Supp. Fig. \ref{CuGeO$_3$_structure}) is a model edge-sharing cuprate well-known for his spin-Peierls transition near 14K \cite{Hase1993}. It is composed by chains of Copper-Oxygen octahedral structures whose direction corresponds to the c-axis of the crystal. In this study, we have used a 100 $\mu$m thick sample of single crystal CuGeO$_3$ (provided by A. Revcolevschi) that we have investigated through midIR-pump Vis-probe spectroscopy on its c-axis (at 8 K and 300 K) and on its b-axis (only 300 K). The results along the c-axis are presented in the main article and few ones about the b-axis are presented in this supplementary. Moreover no strong signature of the spin-Peierls transition has been observed.

\begin{figure}[h!]
\centering
\includegraphics[scale=0.7]{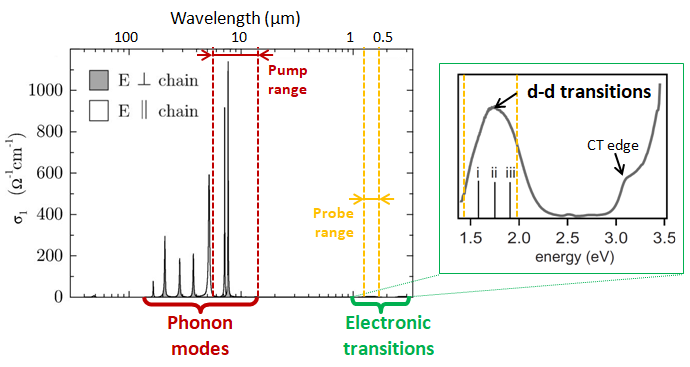}
\caption{Left panel: optical conductivity of CuGeO$_3$ for two polarizations: along the b-axis (perpendicular to the chain) or the c-axis (parallel to the chain). The accessible pump (resp. probe) wavelength range is indicated by red (resp. orange) dashed line. Right panel: zoom on the electronic transitions which shows that the d-d transitions are split of about 0.7 eV from the charge transfer (CT) edge or higher electronic transitions. Left figure adapted from \cite{Damascelli2000} and right figure adapted from \cite{Giannetti2009}.}\label{CuGeO$_3$_optical_properties}
\end{figure}

\noindent For the purpose of this study, it is also interesting to detail the optical properties of CuGeO$_3$. Indeed, it owns many intense phonon modes at low energy \cite{Damascelli2000,Popovic1995} (Supplementary Fig.\ref{CuGeO$_3$_optical_properties}-left) and a group of three phonon-assisted d-d transitions that is isolated in energy from other electronic transitions (Supp. Fig. \ref{CuGeO$_3$_optical_properties}-right). These properties are suitable if one wants to excite specific low energy modes and to probe the response of the three d-d transitions, located between 1.5 and 2 eV, independently from the response of other electronic transitions.

\subsection{Experimental method}
\label{experimental_method}
\begin{figure}[h!]
\centering
\includegraphics[scale=0.6]{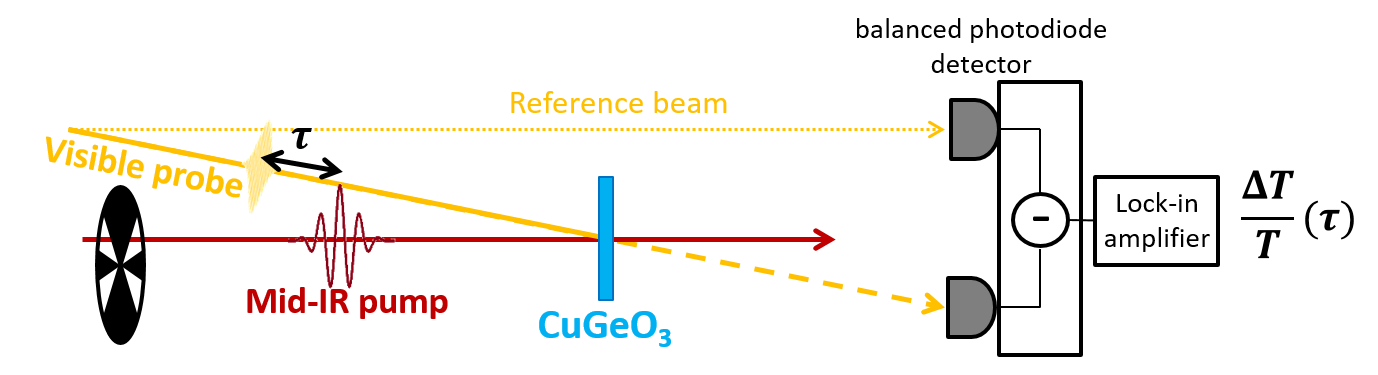}
\caption{Simple scheme of the experimental setup used to measure the transient transmissivity with an accuracy down to 10$^{-5}$.}\label{Experimental_scheme}
\end{figure}

\subparagraph{Description of the setup.} The experiment was performed on our recently developed midIR-pump and visible-probe setup operating up to a repetition rate of 50 kHz. The simple scheme of the experiment is depicted in Supp. Fig.\ref{Experimental_scheme} but more details about it can be found in \cite{Randi2016} and \cite{Giusti2018}. 

\noindent Briefly, the midIR pulses are generated thanks to a Difference-Frequency Generation (DFG) system seeded by a twin Optical Parameric Amplifier (TOPAS, Light Conversion). The pump central wavelength can be tuned from 4 $\mu$m up to 18 $\mu$m keeping an energy bandwidth of about $7 \pm 1$ meV, which corresponds to Fourier-transform limited pulse duration of $260 \pm 40$ fs. The pump pulses are then focused on the sample on a spot size of about $150 \pm 50 \mu$m diameter, which allows to reach fluences up to few mJ.cm$^{-2}$. The pump fluence was not a limitation in this experiment and we have performed some preliminary measurements that have demonstrated the linearity of the observed effects as a function of the pump fluence (not shown here). On the probe arm, a non-collinear OPA (Orphenus-N, Light Conversion) generates visible probe pulses whose central wavelength is tunable from 650 nm (1.91 eV) up to 950 nm (1.30 eV) keeping a bandwidth of about $60 \pm 10$ meV,  which corresponds to Fourier-transform limited pulse duration of $30 \pm 5$ fs (checked with a FROG \cite{Trebino1997}). The probe beam goes toward a delay line and a 90/10 beamsplitter in order to obtain a reference probe beam (10 \%) and a main part (90 \%) that goes through a $\lambda/2$+polarizer device, in order to control its power. The main part is then focused on a spot size of about $75 \pm 25 \mu$m diameter allowing to reach intensities up to few tens of $\mu$J.cm$^{-2}$. For the measurements, an intensity not higher than 2 $\mu$J.cm$^{-2}$ was chosen in order to remain in the probe linear regime. Pump and probe pulses interact in a non-collinear scheme (angle $< 15^{\circ}$) on the CuGeO$_3$ sample (100-$\mu$m thick) that is inside a ultra low vibration closed cycle cryostat (CS204-DMX20-OM, Advanced Research Systems) mounted on 3-axis positioning stage. Moreover both pump and probe beams have a vertical polarization. For each delay step (of 12 fs), we measure the transmitted probe beam with one channel of the balanced photodiode detector and the other channel is used to measure the reference probe beam. The transmissivity ($\Delta$T/T) measurements are carried out by subtracting a reference probe beam signal to the sample-transmitted one in order to remove the intensity fluctuations of the laser. Then, the differential signal is processed by a lock-in amplifier, in phase with an optical chopper wheel (500 Hz) located on the pump arm. The lock-in signal is averaged over 300 ms for every delay step and a complete measurement corresponds to an average of at least 3 scans (5 for most of them). All together, this setup allows to get variations of transmissivity down to 10$^{-5}$. 

subsection{Analysis}

\subparagraph{Zero delay shifts.} 
Due to technical details related with the NOPA-design, it was not possible to keep a constant zero delay while changing the probe wavelength. Thus, for each probe wavelength, the zero delay has been fixed in a post-measurement treatment. In particular, it has been chosen as the starting point of the dynamics, i.e. the beginning of the decreasing or increasing edge of $\Delta$T/T. This choice is justified if we assume that the mechanism leading to the decrease or the increase of $\Delta$T/T is ``suddenly'' triggered by the pump pulse, in other words, if there is no delay between the variation of $\Delta$T/T and the true excitation moment. If so, the $\Delta$T/T decreasing (resp. increasing) edge is fixed by the cross-correlation duration between the pump and the probe. In our case, the probe duration is much shorter than the pump duration (30 fs comparing to 260 fs), therefore the $\Delta$T/T dynamics edge corresponds to the delay when both pulses start to interact which is then chosen as the ``zero delay''.

\begin{figure}[h!]
\centering
\includegraphics[scale=0.5]{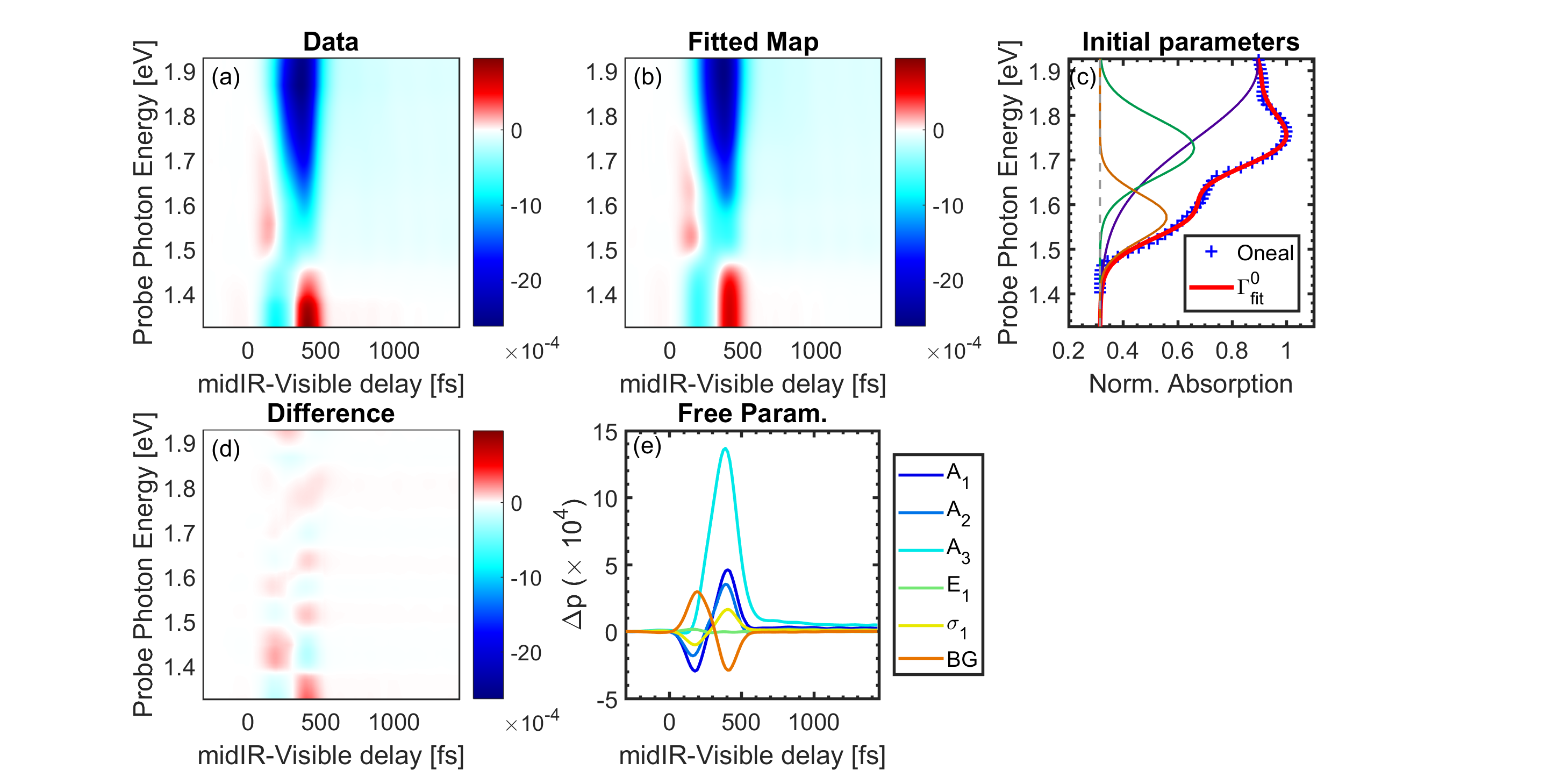}
\caption{(a) Measured data same as in the main article Fig.2. (b) Retrieved fitted map obtained by varying the extracted parameters of the (c) fit with three Gaussians of the optical absorption (adapted from \cite{ONeal2017}. (d) Difference between the data and the fitted map in order to appreciate the quality of the fit. (e) Variation of the free parameters used in (b).}\label{Fitting_procedure}
\end{figure}

\subparagraph{Fitting procedures\label{para_fitting_procedure} .}
In order to make the transmissivity map more intelligible and in relation with a dynamical shaping of the three d-d transitions, we have performed a fit of the measured response in the energy-domain and we have repeated this fit for each time delay. In particular, we have firstly fitted the optical absorption of the d-d transitions, which has been measured by O'Neal et al. \cite{ONeal2017}, by three Gaussians plus a background in order to obtain a set of initial parameters: 
\begin{equation}
\Gamma_{fit}^0(E) = \sum_i A_i^0exp(-(E-E_i^0)^2/[\sigma_i^0]^2) + BG
\end{equation}

\noindent where $A_i^0$, $E_i^0$ and $\sigma_i^0$ respectively correspond to the amplitude, central energy and bandwidth of Gaussian representing the i$^{th}$ d-d transition, and $BG$ accounts for the background absorption, which is close to the value that is observed out of the d-d features (see Supp. Fig. \ref{Fitting_procedure}).
 
\noindent We want to emphasize that we have tried other distributions to fit this optical absorption, especially by using Lorentzian shapes or Fano profiles. However, using Gaussian distributions seems to be the most reproducible and stable manner of fitting the data, i.e. the less sensitive one to initial guess of the parameters. Note also that it might be probable that a non-trivial kind of distribution could correspond to the absorption shape of these phonon-assisted transitions. Indeed, we have shown, through a minimalist model (see \ref{Theory_general_case} Supp. Fig.\ref{ABC_dependency}), that the distribution shape could rather be similar to a ``full and displaced" Maxwell-Boltzmann distribution\footnote{By ``full'', we want to include the positive and negative part of a typical Maxwell-Boltzmann distribution of type $p(E) \propto \frac{E^2}{\sigma_E ^3} exp(-\frac{E^2}{2\sigma_E^2})$. By ``displaced", we mean that the energy axis has to be shifted in such a way that E is replaced by $E'=E-\epsilon$ in the previous formula.} whose central energy is the one of the d-d transition. Obviously, this kind of distribution has the drawback to be hard to interpret whereas the Gaussian distribution parameters are easily intelligible.
Therefore, we use the extracted parameters from the Gaussian fits of the linear response as input parameters to construct a fitting function for the transient transmissivity map. This fitting function is defined by:

\begin{equation}
\left(\frac{\Delta T}{T} (\tau) \right)_{fit}  =10^{(\Gamma_{fit}^0(E)-\Gamma_{fit}(\tau,E))}-1
\end{equation}

\noindent where $\tau$ is the pump-probe delay and each Gaussian parameter ($A_i$, $E_i$ or $\sigma_i$) can be free to change with the delay or fixed to the initial values ($A_i^0$,$E_i^0$,$\sigma_i^0$). We tested several couples of a free parameters to fit the measurement with the aim to find the best compromise among (i) obtaining a good fitted map (obviously done by letting all the parameters free) and (ii) using a minimum number of free parameters. After different tests, we ended up with a fitting function that only requires the three amplitudes ($A_i(\tau)$), the central energy and bandwidth of the first d-d transition ($E_1(\tau)$, $\sigma_1(\tau)$) an the background ($BG(\tau)$) to be free in order to obtain a good agreement with the measured map (see Supp. Fig. \ref{Fitting_procedure}). The evolution of each parameter is plotted in Supp. Fig. \ref{Fitting_procedure}.e and we can observe that it is possible to differentiate the response of each transition since the variation of the amplitudes are dynamically specific. Moreover, for the first d-d transition, the best way to fit the observed transient transparency is to let free the central energy and bandwidth of the transition ($E_1$ and $\sigma_1$). Moreover the background constant ($BG$) has to evolve in order to reproduce the low energy features (below 1.45 eV). We interpret this as if an ingredient was missing in the chosen distribution used to fit the linear response. Indeed, we could imagine that a realistic distribution could have some contributions in an energy range out of the measured features of the d-d transitions. Even if this point remains unclear, we can however claim that the lowest energy observed dynamics is induced by the coherent midIR excitation since it is not only negative as it was the case in previous studies \cite{Giannetti2009,Yuasa2008}.

\section{Complementary results and discussion}

\subsection{Thermal effects}
\label{thermal_effects}

\subparagraph{Effects of the initial sample temperature.} The CuGeO$_3$ d-d transitions have a strong dependence as a function of the temperature. In particular, the d-d absorption amplitude increases when the temperature increases and we can justify this behavior thanks to the developed theoretical model (see. section \ref{theory}). Besides, all the d-d transitions shift toward lower energy for an higher temperature (see Supplementary Fig.\ref{Map8K_vs_Map300K}.a) and they also broaden in the linear response case. Therefore, we also wanted to study the effects of the initial sample temperature on the transient response of the d-d transitions. These results are shown in Supplementary Fig.\ref{Map8K_vs_Map300K}.b-c, where we compare two transient maps measured at 8 K and 300 K, in the same pump and probe conditions ($\lambda_{pump} = 9 \mu$m). We can observe that, at 300 K, the transient signal is very similar (in time and amplitude) to the one at 8 K but it has been shifted toward lower energy. We have also plotted some colored lines that indicate the central energy of the fitted Gaussians in both cases. The energy shift of the overall map is similar to the one that is observable in the linear absorption response case (about 60 meV), which supports the idea of midIR-induced distortions that would similarly impact the d-d transitions whatever the temperature is. It notably opens the perspectives in providing a way to induce some specific electronic properties at room temperature. Finally, note that CuGeO$_3$ has a Spin-Peierls transition around 14 K but no particular signature of this transition has been observed in this experiment.

\begin{figure}[h!]
\centering
\includegraphics[scale=0.4]{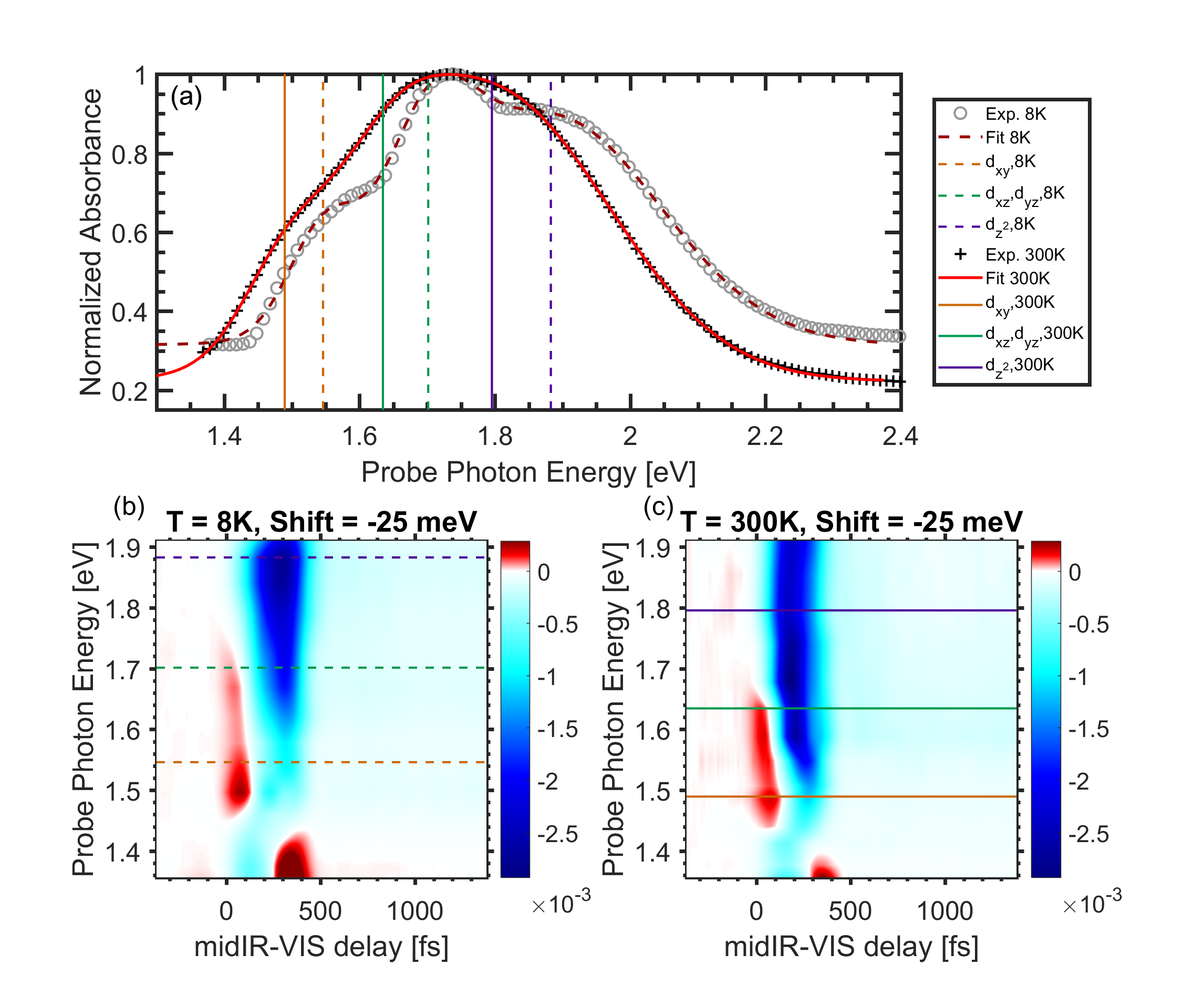}
\caption{(a) Linear absorption measurements adapted from \cite{ONeal2017} at 8 K (gray circle) and 300 K (black cross) and their respective fit (dashed dark red at 8K, plain red at 300K). As a guide for the eyes, we have also plotted the central energy of the 3 fitted Gaussians representing the 3 d-d transitions. (b-c) Transmissivity maps at 8 K and 300 K for the same conditions of pump and probe beams. The indicated shift (-25 meV) corresponds to a the energy shift used to align the data measured by ONeal \textit{et al.} with some calibrations performed with our spectrometer. }\label{Map8K_vs_Map300K}
\end{figure}

\subparagraph{Pump-induced thermal effects} As discussed in the main text, it is important to estimate the possible temperature increase due to the pump excitation in order to understand its potential role in the transmissivity maps. For that purpose, we have considered that the absorbed energy from the pump ($\Delta Q_{pump}$) is transferred into heat, which allows defining the temperature increase as:
\begin{equation}
\delta T = \frac{\Delta Q_{pump}}{C_T \times n_{CuGeO_3}} \, , \, 
\left\{
        \begin{aligned}
        \Delta Q_{pump} = F  S (1- R -T) \, \, \, \, in \, [J] \\
		n_{CuGeO_3} = \frac{S \times L}{V_{cell}\times N_A}  \, \, \, \, in  \,[mol]
     \end{aligned}
   \right.
\end{equation} 

\noindent where $F$ is the midIR fluence (1 mJ.cm$^{-2}$), $S$ is the focus area (disk of radium $r=75 \mu$m), $R$ is the reflection on the surface ($\simeq 6.5\%$ at $9 \mu$m \cite{Damascelli2000}), $T$ is the transmission coefficient (see below, eq.\ref{T_formula}), $L$ is the sample thickness ($100 \mu$m), $V_{cell}$ is the volume of one CuGeO$_3$ unit cell ($59.9 \si{\angstrom}^3$), $N_{A}$ is the Avogadro constant and $C_T$ is the heat capacity ($0.43$ J.K$^{-1}$.mol$^{-1}$ at 8 K \cite{Liu1995}, and $100$ J.K$^{-1}$.mol$^{-1}$ at 300 K\cite{Weiden1995}). To compute $\Delta Q_{pump}$, one needs the transmission coefficient which can be measured (not done in this study) or computed by the following standard formula \cite{Zeman1999} :
\begin{equation}
T = \frac{(1-R)^2exp(-\alpha_\lambda L_{sample})}{1+R^2exp(-2\alpha_\lambda L_{sample})}
\label{T_formula}
\end{equation}
where $\alpha_{\lambda}$ is the absorption coefficient which depends on the wavelength. Then, we base our reasoning on the measurements of $R$ and $T$ done by Damascelli et al. \cite{Damascelli2000} on a very broad spectral range. They have found $R=0.065$ and $T=0.78$ at $\lambda = 9 \mu$m which gives, thanks to eq.\ref{T_formula}, $\alpha_{9\mu m} L_{sample} \simeq 0.11$. As they did not specify their sample thickness, we had to retrieved it by using their measurements, done one the same sample, at $\lambda = 730 nm = 1.7 eV$ ($R=0.114$ and $T=0.016$). This wavelength corresponds to the d-d band whose absorption coefficient has been previously reported to lay between 200 and 600 cm$^{-1}$ (depending on the source \cite{ONeal2017,Bassi1996}). Therefore, we can retrieve $L_{sample}$ that has been used by Damascelli et al. (195 $\mu$m to 65 $\mu$m) and then estimate $\alpha_{9\mu m}$: between 5.6 cm$^{-1}$ and 16.9 cm$^{-1}$. This last values permit to get $\Delta Q_{pump}$ and finally to estimate a range of pump-induced temperature increase $\delta T$: between 0.94 K and 1.66 K for an initial temperature of 8 K and between 4 mK and 7.3 mK for an initial temperature of 300 K.

\noindent According to this estimation, the temperature increase, linked to a complete transfer into heat of the absorbed pump energy, is about 2 orders of magnitude higher at 8 K than the one at 300 K. Nevertheless, we have observed that the transmissivity maps at these two sample temperatures are very similar: we only observe a shift of the overall map toward lower energy (see Supplementary Fig.\ref{Map8K_vs_Map300K}). In particular, this shift seems to be linked with the difference in the initial sample temperature which induces a overall shift of the d-d band features toward lower energies of about 60 meV. In other word, the pump-induced thermal effects are certainly negligible or not visible on the probed timescale with respect to the effect of the initial sample temperature. Moreover, the developed model results (see Supplementary Fig.\ref{T_B_dependcies_300K_01} for T = 300 K), are also in favor of this interpretation. These theoretical results show that the absorption spectral distribution variations (central energy and the energy bandwidth shifts) which are induced by a $\delta T = 10^{-2}$ K at 300 K (resp. a $\delta T = 1$ K at 8 K) are negligible with respect to the ones induced by a relevant displacement variation ($\Delta B_t$) in the same conditions.

\subsection{Pump wavelength dependency}
\label{pump_wavelength_dependency}
\begin{figure}[h!]
\centering
\includegraphics[scale=0.31]{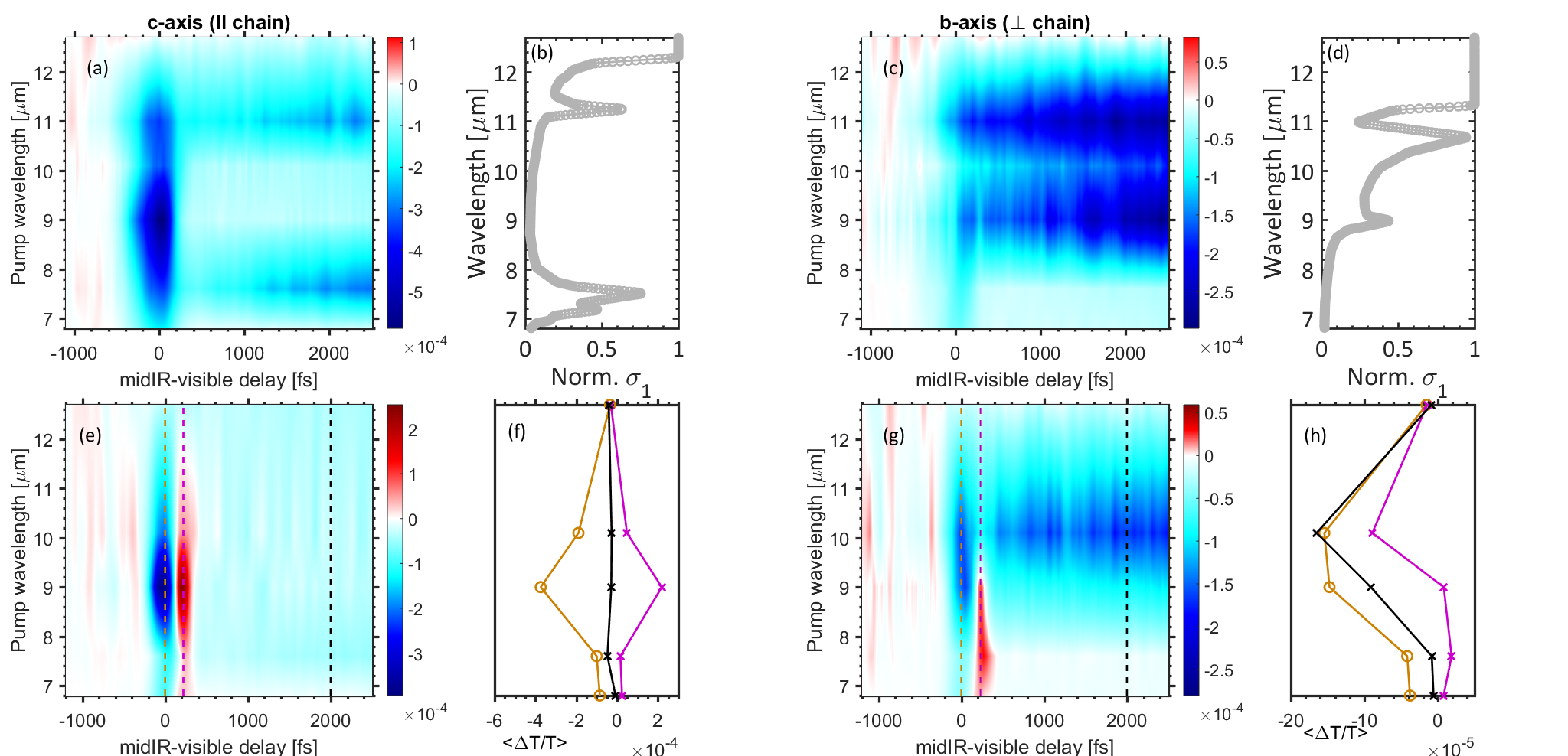}
\caption{Top (a-d): Comparison of the pump-wavelength dependence transmissivity maps for (a) the c-axis and (c) the b-axis in the case of $E_{probe}=1.7eV$. For each crystallographic axis, the optical conductivity has been plotted on the same energy range: (b) $\sigma_{1,c-axis}$ and (d) $\sigma_{1,b-axis}$ adapted from \cite{Damascelli2000}. Bottom (e-h): Comparison of pump-wavelength dependence transmissivity maps for (e) the c-axis and (g) the b-axis in the case of $E_{probe}=1.35eV$. The different dashed lines correspond to the cut at specific delays: 0 fs (orange), 275 fs (magenta), 2 ps (black), which are represented in the right panel of each map: (f) $\Delta T/T_{c-axis}$ and (h)  $\Delta T/T_{b-axis}$.}\label{Comparison_c-axis_b-axis}
\end{figure}

In order to obtain a better insight on the role of the pump for probed dynamics, we have performed some preliminary transmissivity measurements along the c-axis and the b-axis of CuGeO$_3$, at 300 K, as a function of the pump wavelength and for two given probe photon energies (Supp. Fig.\ref{Comparison_c-axis_b-axis}). Besides, in the current subsection, the data are issued from an anterior set of measurements for which the pump duration and the probe duration were longer: it justifies the mismatches comparing to the data that are shown in the main text (Fig.2). 

\subparagraph{Pump wavelength dependence at E$_{probe} = 1.7$, c-axis vs b-axis (Supplementary Fig.\ref{Comparison_c-axis_b-axis}.a-d).}
Compared to the c-axis (main text, Left Panel Fig.2), the b-axis does not show a quick variation of $\Delta$T/T around delay zero but only long timescale population dynamics are observed. These slow population dynamics are especially intense for two pump wavelengths around 9 $\mu$m and 11 $\mu$m. Such as for the c-axis, these long timescale dynamics appear for pump wavelengths which correspond to peaks in the optical conductivity measurement (Supplementary Fig.\ref{Comparison_c-axis_b-axis}.d), that are around 9 $\mu$m and 10.7 $\mu$m for  the b-axis optical conductivity.
Besides, for both axis, some measurements have been performed for pump wavelength from 13 $\mu$m to 17 $\mu$m, but nothing clear (very noisy signals) was observed. This is is notably justified by the fact that the midIR light cannot propagate inside the media at these wavelengths (see \ref{simulation_finite_element_subsection}).

\subparagraph{Pump wavelength dependence at E$_{probe} = 1.35$, c-axis vs b-axis (Supplementary Fig.\ref{Comparison_c-axis_b-axis}.e-h).}
We also wanted to understand better the role of the pump concerning the transient transparency that has been observed in the low-energy range (below 1.45 eV), namely out of the d-d transitions. Therefore, we have performed a set of pump wavelength dependent measurements keeping the probe photon energy at 1.35 eV and the results are shown in Supp. Fig.\ref{Comparison_c-axis_b-axis}.e-h. We can clearly observe that, on the c-axis, the maximum of this transient transparency is around $\lambda_{pump} \approx 9 \mu$m whereas on the b-axis a maximum of this transparency appears around $\lambda_{pump} \approx 8 \mu$m. The c-axis map clearly confirms that the maximum amplitude of the coherent effects is obtained for $\lambda_{pump} = 8 \mu$m which is located out of the phonon modes.

\begin{figure}[h!]
\centering
\includegraphics[scale=0.35]{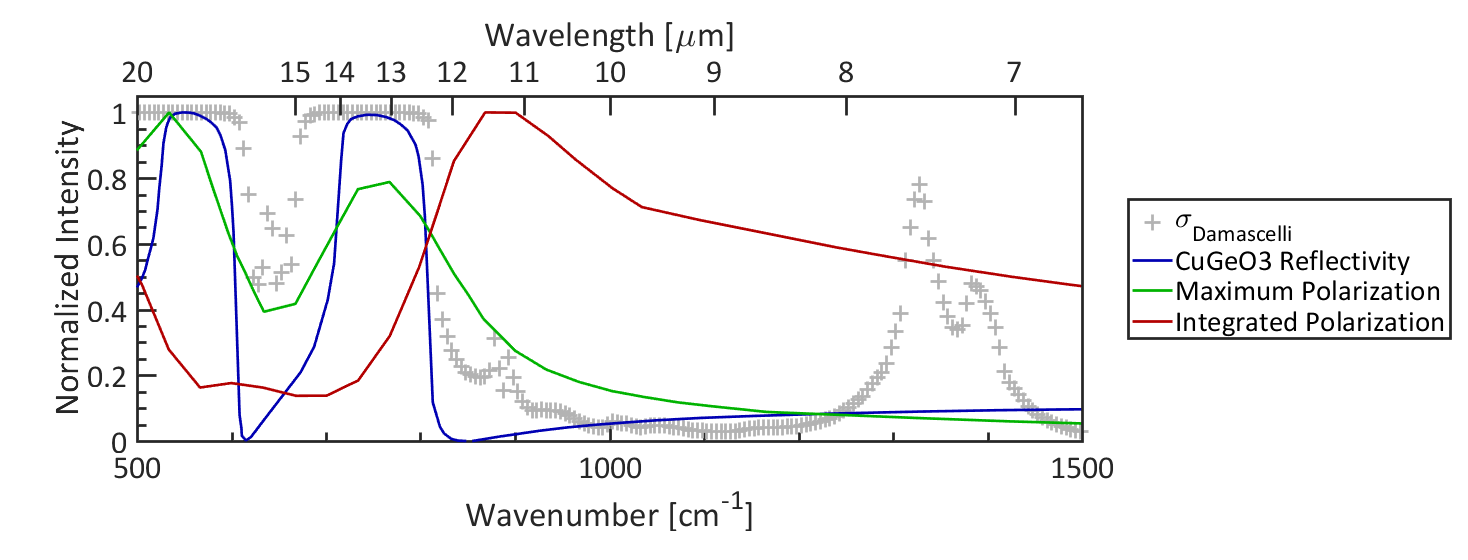}
\caption{Computed polarization vector maximum for CuGeO$_3$ (blue curve is the reflectivity) at the surface (green curve) or taking into account the propagation effect inside (red curve).}\label{simulation_finite_element}
\end{figure}

\subparagraph{Finite difference time domain (FDTD) simulation of electromagnetic wave propagation}
\label{simulation_finite_element_subsection}
The lattice polarization (P(x,t)) induced by a pump pulse has been computed using a home-made FDTD code \cite{Cartella2017,Taflove2005}. First, we have calculated the maximum polarization amplitude that develops inside the material as a function of pump wavelength considering the CuGeO$_3$ measured reflectivity. Typically, this maximum is reached at the sample surface and at time zero, i.e. when the pulse hits the material. The result of this calculation is displayed in Supp. Fig.\ref{simulation_finite_element} for a pump pulse duration of 200 fs (green curve). As expected the maximum value is reached for photon energies within the reststrahlen band(s), i.e. around 750 cm$^{-1}$ ($\approx 13.5 \mu$m) and 530 cm$^{-1}$ ($\approx 19 \mu$m), which is where the screening is more efficient. Secondly, we have estimated the effects of penetration dept mismatch between the pump and the probe in a transmission experiment. Intuitively, a pump pulse, which is tuned off-resonantly with respect to a phonon mode, will penetrate more into the material. Therefore, even if this pulse has a lower polarization intensity than one which is tuned in-resonance, the total polarization effects all along the sample might be higher. In first approximation, this quantity can be estimated by calculating the integrated polarization in space for a given delay and then by taking the maximum value of the resulting vector. Thus, the integrated polarization takes into account the propagation effects inside the material as it is depicted in Supp. Fig.\ref{simulation_finite_element} (red curve). We can clearly observe that the integrated polarization peaks around 900 cm$^{-1}$ ($\approx 11 \mu$m), on the right side of the reststrahlen band. This value is not so far from the one that was used during the experiment (9 $\mu$m). Many factors could be considered to get results that are more realistic: (i) the group velocity of the probe since integrating in space for a given delay is equivalent to having a probe with infinite velocity, (ii) the nonlinear response of the lattice which is driven to large amplitudes and (iii) absorption peaks at higher energies. Above all, the observed blue shift trend is completely compatible with the measurements shown in Supplementary Fig.\ref{Comparison_c-axis_b-axis}, where higher coherent effects are observed for a pump wavelength that is also blueshifted with respect to the targeted phonon mode.

\subsection{Phonon mode on long timescale}
\label{longtime}

The zoom in the transmissivity map at 8 K shows an additional feature on long timescale: a phonon mode is excited (see Supplementary Fig.\ref{Zoom_Map8K} for 8K). The extracted frequency is 182 cm$^{-1}$ which is thus associated to the $A_g$ phonon mode at 187 cm$^{-1}$ \cite{Popovic1995}. This demonstrates the possibility to excite Raman modes through anharmonic couplings on long timescale.

\begin{figure}[h!]
\centering
\includegraphics[scale=0.36]{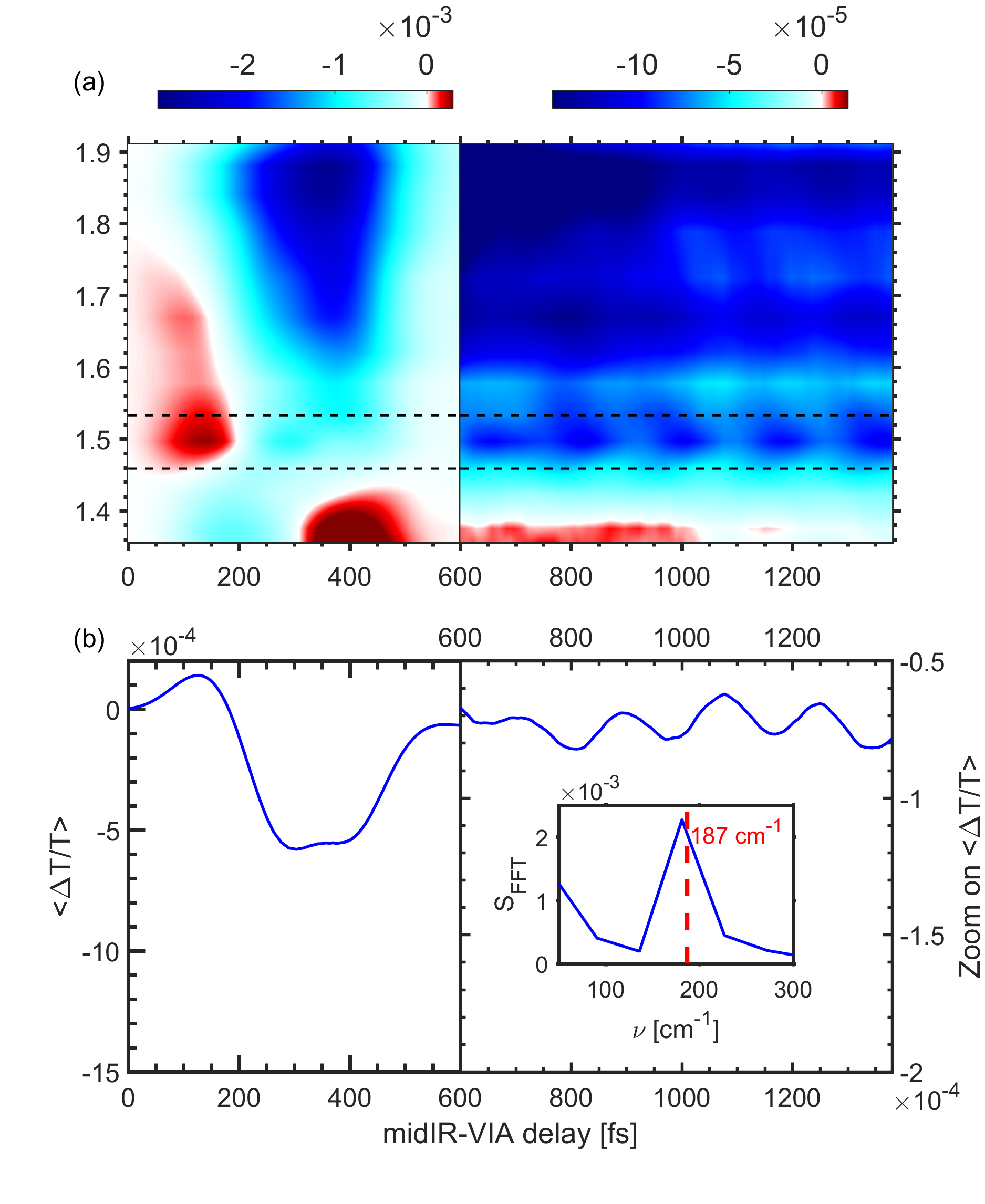}
\caption{(a) Right part: zoom in transmissivity map at 8 K, the color scale has been divided by 20. (b) Right part: zoom on $<\Delta T/T>$  that has been averaged between 1.46 and 1.53 eV (dashed lines on (a)). The inner panel shows the Fourier transform of the right panel signal and the red dashed line corresponds to the lowest $A_g$ phonon mode at 187 cm$^{-1}$.}\label{Zoom_Map8K}
\end{figure}

\newpage
\section{Theory}
\label{theory}
In the following, we discuss the details about the theoretical model used in the main text. A localized d-d transition is described by a two-level system interacting with a single vibrational mode of the crystal represented by a quantum harmonic oscillator. The probe light is first described as a classical field in Section \ref{equi} while in Section \ref{Theory_general_case} we adopt a fully quantum picture. In this context, by choosing a suitable Hamiltonian for the electron-phonon-photon interaction we can compute the average number of transmitted photons at a certain frequency, up to leading order in perturbation theory. This in turn gives information about the absorption spectrum of the sample in the frequency range of the dressed electronic transition.
The calculation is performed firstly by considering the sample in equilibrium and afterwards by taking into account the displacement induced by the pump pulse.
The expected behavior in temperature, consistent with the experimental findings \cite{ONeal2017}, is correctly predicted by our calculations. A overall enhancement of the integrated absorption is predicted as well, together with a shift of the average frequency of the transmitted photon distribution.
Finally, in Section \ref{molecule} a justification of the model is given in terms of standard theoretical treatment of molecular spectroscopy which is well suited for this kind of localized transitions.

\subsection{Phonon-dressed d-d transition}

The equilibrium situation is described through the following model Hamiltonian for the phonon-electron interaction
\begin{equation}
H_{ph-el} \equiv H =\omega \, b^\dag b + \epsilon \, d^\dag d + M d^\dag d \, (b + b^\dag),
\end{equation}
where $b, b^\dag$ are bosonic operators describing the vibrational degree of freedom and $d, d^\dag$ are fermionic operators describing the electronic transition. The parameters $\omega$ and $\epsilon$ represent the bare phonon frequency and electronic transition energy, respectively, while $M$ is the coupling between the two degrees of freedom.
 
\noindent This model Hamiltonian can be explicitly diagonalized \cite{Mahan2000}. Indeed, one can find a diagonal Hamiltonian $\widetilde{H}$
\begin{equation}
\widetilde{H} = \omega \, b^\dag b + \left( \epsilon - \frac{M^2}{\omega} \right) d^\dag d
\end{equation}
that is related to $H$ by a unitary transformation $U$
\begin{equation}
\widetilde{H} = U^\dag \, H \, U, \quad U = \mathrm{e}^{-\frac{M}{\omega} d^\dag d (b^\dag - b)}.
\end{equation}
As a consequence, the eigenvalues of $\widetilde{H}$ correspond to the eigenvalues of $H$
\begin{equation}
sp(H) = \Big\{n \omega \,\Big| n \in \mathbb{N}\Big\} \cup \Big\{n \omega + \epsilon - \frac{M^2}{\omega} \,\Big| n \in \mathbb{N}\Big\},
\end{equation}
while the eigenvectors $| \psi \rangle_{n}^{m}$ of $H$ are obtained from the eigenvectors $| \widetilde{\psi} \rangle_{n}^{m}$ of $\widetilde{H}$ through the unitary operator $U$
\begin{equation}
| \psi \rangle_{n}^{m} = U | \widetilde{\psi} \rangle_{n}^{m} , \quad | \widetilde{\psi} \rangle_{n}^{m} = |n\rangle \otimes |m\rangle = \frac{(b^\dag)^n}{\sqrt{n !}} |0\rangle \otimes (d^\dag)^m |0\rangle,
\end{equation}
with $n \in \mathbb{N}$ and $m \in \{ 0,1\}$. More explicitly, the action of $U$ has no effect on the eigenstates $| \widetilde{\psi} \rangle_{n}^{0}$ corresponding to the electronic ground state, namely $ | \psi \rangle_{n}^{0}  =| \widetilde{\psi} \rangle_{n}^{0}$, while the eigenstates describing the electronic excited state are displaced proportionally to the coupling constant $M$
\begin{equation}
| \psi \rangle_{n}^{1} = \mathrm{e}^{-\frac{M}{\omega} (b^\dag - b)} |n\rangle \otimes |1\rangle .
\end{equation}

\subsection{Probe-target interaction (equilibrium)}
\label{equi}

In the pump-probe setup of our experiment two different light pulses interact with the sample. An infrared pump pulse induces coherent vibrations in the crystal (along a specific normal mode) and after a delay-time $t$ a visible probe pulse induces electronic transitions. 

\noindent Let us concentrate for the moment on the interaction between the probe light and the electron-phonon system. We start describing the pulse as a classical field $E$ and assuming the sample system to be in the electronic ground state with a thermal distribution of vibrational excitations. Using the notation introduced in the previous section, the initial state for the electron-phonon system at a given inverse temperature $\beta$ reads
\begin{equation}\label{thermstate}
\varrho = \varrho_\beta \otimes | 0 \rangle\langle 0 |, \quad \varrho_\beta = \frac{\mathrm{e}^{-\beta \omega n}}{\mathrm{Tr}(\mathrm{e}^{-\beta\omega b^\dag b})} |n \rangle\langle n |.
\end{equation}

\noindent According to the Fermi Golden Rule we expect the absorption spectrum to be of the form
\begin{equation}
A(\nu)= \sum_\ell \Gamma_\ell \,\delta(\nu - \epsilon + M^2/\omega -\omega \ell)
\end{equation}
where the quantities $\Gamma_\ell$ are related to the transition probability rates induced by the dipole moment operator $D$ from the electronic ground state to the electronic excited state, producing $\ell$ phonons. In particular, considering a thermal initial state \eqref{thermstate} one has for $\Gamma_\ell$ the following expression
\begin{equation}
\Gamma_\ell= \sum_n \frac{\mathrm{e}^{-\beta\omega n}}{Z_{\beta}} |\langle n+\ell |\otimes \langle 1 | \mathrm{e}^{\frac{M}{\omega} (b^\dag - b)} \,D\, | n \rangle \otimes | 0 \rangle|^2 .
\end{equation}
In general, given some coupling strength $C$, we could expect the dipole moment operator to be of the form $\mu= C ( d+d^\dag)$ because this operator allows transitions from the electronic ground state to the excited state. In our setting, due to the electron-phonon coupling, we could think the coefficient $C$ to be indeed phonon-dependent and, in particular, to be of the form $\mu_0 (b+b^\dag)$. This choice will be justified in the following by means of standard molecular physics arguments.

\noindent With these assumptions, the overall absorption coefficient at finite temperature reads 
\begin{align}\label{abso1}
\Gamma &= \int \mathrm{d}\nu A(\nu) =\alpha^2\mu_0^2 \sum_{m}\sum_n \frac{\mathrm{e}^{-\beta \omega n}}{Z_\beta} |\langle m | \mathrm{e}^{\frac{M}{\omega} (b^\dag - b)} (b+b^{\dag}) | n \rangle |^2 \nonumber\\
&= \alpha^2\mu_0^2 \mathrm{Tr}\left[ \varrho_\beta (2 b^\dag b + 1)   \right] = \alpha^2\mu_0^2 \coth\left(\frac{\beta\omega}{2} \right).
\end{align}
where $\alpha^2$ is proportional to the intensity of the field. For simplicity, we use $\alpha^2=1$ in all the following developments of the model. The expression of $\Gamma$ is in agreement with experimental findings \cite{ONeal2017,Bassi1996} and previous theoretical studies \cite{Ballhausen1962}.
The computation of the single absorption lines is a bit more involved. Explicitly, the quantity to be determined is the amplitude of the absorption line corresponding to the transition energy $\Delta E (\ell)= \epsilon-\frac{M^2}{\omega} + \omega \ell$, namely
\begin{equation}\label{speclines}
\Gamma_\ell = \mu_0^2 \sum_n \frac{\mathrm{e}^{-\beta\omega n}}{Z_{\beta}} |\langle n+\ell | \mathrm{e}^{\frac{M}{\omega} (b^\dag - b)} (b+b^{\dag}) | n \rangle |^2 .
\end{equation}
The first step is the computation of the matrix element $\langle m | \mathrm{e}^{\frac{M}{\omega} (b^\dag - b)}  | n \rangle$. Using the following algebraic property
\begin{equation}
\mathrm{e}^{\frac{M}{\omega} (b^\dag - b)} = \mathrm{e}^{\frac{M}{\omega} b^\dag } \mathrm{e}^{-\frac{M}{\omega} b } \mathrm{e}^{-\frac{M^2}{2\omega^2}  },
\end{equation} 
one can rewrite the matrix element in a convenient way (for $n \geq m$)
\begin{align}\label{matel1}
\langle m | \mathrm{e}^{\frac{M}{\omega} (b^\dag - b)}  | n \rangle &= \mathrm{e}^{-\frac{M^2}{2\omega^2}} \sum_{j=0}^m \sum_{i=0}^n (-1)^{i} \left(\frac{M}{\omega}\right)^{i+j}\frac{1}{i! j!}\frac{\sqrt{n! m!}}{\sqrt{(n-i)! (m-j)!}} \langle m-j | n-i \rangle \nonumber \\
&= \mathrm{e}^{-\frac{M^2}{2\omega^2}} \sum_{j=0}^m (-1)^{n-m+j}\left(\frac{M}{\omega}\right)^{n-m+2j} \frac{\sqrt{n! \, m!}}{(n-m+j)!\, j! \,(m-j)!} = \nonumber \\
&= \mathrm{e}^{-\frac{M^2}{2\omega^2}} (-1)^{n-m} \left(\frac{M}{\omega}\right)^{n-m} \frac{\sqrt{m!}}{\sqrt{n!}} L_{m}^{n-m}\left(\frac{M^2}{\omega^2}\right),
\end{align}
where the generalized Laguerre polynomials $L_{i}^{j}(x)$ are defined as follows \cite{Gradsteyn2007}
\begin{equation}
L_{i}^{j}(x) = \sum_{t=0}^i \frac{(-1)^{t}}{t!}x^t \frac{(i+j)!}{(j+t)! \, (i-t)! }.
\end{equation}
Also, for $n \leq m$ one can use $\langle m | X | n \rangle = (\langle n | X^\dag | m \rangle )^*$ and arrive at
\begin{equation}\label{matel2}
\langle m | \mathrm{e}^{\frac{M}{\omega} (b^\dag - b)}  | n \rangle = \mathrm{e}^{-\frac{M^2}{2\omega^2}} \left(\frac{M}{\omega}\right)^{m-n} \frac{\sqrt{n!}}{\sqrt{m!}} L_{n}^{m-n}\left(\frac{M^2}{\omega^2}\right).
\end{equation}
Coming back to equation \eqref{speclines} one can see that the action of $b^\dag + b$ gives two matrix elements of the kind discussed before, namely
\begin{equation}
\sqrt{n} \,\langle n+\ell | \mathrm{e}^{\frac{M}{\omega} (b^\dag - b)}  | n-1 \rangle + \sqrt{n+1} \, \langle n+\ell | \mathrm{e}^{\frac{M}{\omega} (b^\dag - b)}  | n+1 \rangle,
\end{equation}
that in turn can be rewritten using \eqref{matel2} (assume $\ell \geq 1$ for now)
\begin{align}
&\mathrm{e}^{-\frac{M^2}{2\omega^2}} \left[ \left(\frac{M}{\omega}\right)^{\ell+1} \frac{\sqrt{n!}}{\sqrt{(n+\ell)!}} L_{n-1}^{\ell+1}\left(\frac{M^2}{\omega^2}\right)+ (n+1)\left(\frac{M}{\omega}\right)^{\ell-1} \frac{\sqrt{n!}}{\sqrt{(n+\ell)!}} L_{n+1}^{\ell-1}\left(\frac{M^2}{\omega^2}\right) \right] = \nonumber \\
&= \mathrm{e}^{-\frac{M^2}{2\omega^2}} \left(\frac{M}{\omega}\right)^{\ell-1} \frac{\sqrt{n!}}{\sqrt{(n+\ell)!}} \left[ \frac{M^2}{\omega^2}L_{n-1}^{\ell+1}\left(\frac{M^2}{\omega^2}\right) + (n+1)L_{n+1}^{\ell-1}\left(\frac{M^2}{\omega^2}\right)  \right].
\end{align}
One can now exploit the recurrence relation of Laguerre polynomials \cite{Gradsteyn2007}
\begin{equation}
(n+1)L_{n+1}^{\ell-1}(x) + x L_{n-1}^{\ell+1} (x) = (\ell-x) L_{n}^{\ell} (x),
\end{equation}
and arrive at
\begin{equation}
\langle n+\ell | \mathrm{e}^{\frac{M}{\omega} (b^\dag - b)} (b+b^\dag)  | n \rangle = \mathrm{e}^{-\frac{M^2}{2\omega^2}} \left(\frac{M}{\omega}\right)^{\ell-1} \frac{\sqrt{n!}}{\sqrt{(n+\ell)!}} \left( \ell- \frac{M^2}{\omega^2}   \right) L_{n}^{\ell} \left(\frac{M^2}{\omega^2}\right).
\end{equation}
Therefore, it remains to compute the quantity
\begin{equation}
\Gamma_\ell = \mu_0^2 \sum_{n=0}^{\infty} \frac{\mathrm{e}^{-\beta\omega n}}{Z_{\beta}} \mathrm{e}^{-\frac{M^2}{\omega^2}} \left(\frac{M^2}{\omega^2}\right)^{\ell-1} \frac{n!}{(n+\ell)!} \left( \ell- \frac{M^2}{\omega^2}   \right)^2 \left( L_{n}^{\ell} \left(\frac{M^2}{\omega^2}\right) \right)^2.
\end{equation}
This can be done by means of the so-called Hardy-Hille formula \cite{Gradsteyn2007}
\begin{equation}
\sum_{n=0}^{\infty} \frac{n!}{(n+\ell)!} t^n L_n^\ell(x) L_n^\ell(y) = \frac{\mathrm{e}^{-\frac{(x+y)t}{1-t}}}{(xyt)^{\ell/2} (1-t)} I_{\ell}\left( \frac{2\sqrt{xyt}}{1-t} \right),
\end{equation}
where $I_\ell(x)$ is a modified Bessel function of the first kind.
The final expression reads
\begin{equation}\label{gammal}
\Gamma_\ell = \mu_0^2 \frac{\omega^2}{M^2}\left( \ell- \frac{M^2}{\omega^2} \right)^2 \mathrm{e}^{\beta\omega \ell/2} \mathrm{e}^{-\frac{M^2}{\omega^2}\coth(\beta\omega/2)} I_\ell\left( \frac{M^2}{\omega^2 \sinh(\beta \omega/2)} \right).
\end{equation}
A similar treatment can be used to study the case $\ell <1$ and it turns out that the expression \eqref{gammal} is true for any $\ell$, using the property $I_{-\ell}= I_{\ell}$. Some information can be extracted by looking at the asymptotic behavior of the modified Bessel function for small or large argument
\begin{equation}
I_\ell(x) \simeq \frac{1}{\ell !} \left(\frac{x}{2}\right)^\ell, \qquad 0<x\ll \sqrt{\ell+1}. \label{asy1} 
\end{equation}

%\end{align}
In particular, when the phonon-electron coupling is small $M\to 0$, so that one can use the relation \eqref{asy1}, it turns out that
\begin{align}
&\Gamma_\ell \simeq \mu_0^2 \ell^2 \frac{\mathrm{e}^{\beta\omega \ell/2}}{|\ell|!} \frac{1}{2\sinh^{|\ell|}(\beta \omega/2) } \left(\frac{M^2}{2\omega^2}\right)^{|\ell|-1} \quad \ell \neq 0,  \\
&\Gamma_0 \simeq \mu_0^2 \frac{M^2}{\omega^2}.
\end{align} 
As a consequence, for vanishing coupling $M$ only the absorption lines with $\ell=1$ or $\ell=-1$ are non-zero and the line with $\ell=-1$ is $\mathrm{e}^{-\beta\omega}$ smaller than the other. In the zero-temperature limit $\beta\to \infty$ only the latter survives (the system is initially in the ground state and cannot lower the number of phonons).

\subsection{Probe-target interaction (general case)}
\label{Theory_general_case}

In the general case, when the initial state is not diagonal in the Hamiltonian eigenbasis because it has been modified by the pump pulse, the Fermi Golden Rule cannot be applied. Therefore, we use here a more general treatment where the probe light is considered explicitly as a quantum field and it is measured after the interaction with the sample.
The interaction between the probe light pulse and the excited sample is described trough the following interaction Hamiltonian
\begin{equation}
H_{int} = \mu_0 p  \sum_k (a^\dag_k + a_k), \quad p= (b+b^\dag)(d+d^\dag),
\end{equation}
where the bosonic operators $a_k$ are related to the light at frequency $\nu_k$.
The evolution of the mean photon number $a^\dag_j a_j$ at a certain frequency $\nu_j$    in a time-interval $\tau$ can be computed with a first order Dyson series 
\begin{align}
&\mathrm{Tr}\Big( \varrho \, U^\dag(\tau) a^\dag_j a_j U(\tau) \Big) \simeq \\ &\mathrm{Tr}\left( \varrho  \left( a^\dag_j a_j + i \lambda \int_0^\tau \Big[H_{int}(s), a^\dag_j a_j\Big]\mathrm{d}s - \lambda^2 \int_0^\tau \mathrm{d}s \int_0^s \mathrm{d}u \Big[ H_{int}(u), \Big[ H_{int}(s), a^\dag_j a_j  \Big]\Big] \right)\right).
\end{align}
The first term $\mathrm{Tr}\left( \varrho \,  a^\dag_j a_j \right)$ is the unperturbed light intensity corresponding to $|\alpha_j|^2$.
By choosing the initial state in the form
\begin{equation}
\varrho = \overline{\varrho} | \alpha \rangle\langle \alpha |, \quad \overline{\varrho}= \sum_\ell p^0_\ell \, | \ell ,0 \rangle \langle \ell, 0 | ,
\end{equation}
where $|\alpha\rangle$ is a multi-photon coherent state, describing the proble pulse, $a_j\vert\alpha\rangle=\alpha_j\vert\alpha\rangle$. One gets a vanishing first order contribution, so that a second order calculation is needed. This choice of the initial state, diagonal in the energy basis, is made to compare this approach to the previous one. More general displaced thermal states will be considered later on. The second order term $\Gamma^{(2)}$ reads
\begin{align}
\Gamma^{(2)} &= \mu_0^2 \Big( 8i \sum_k \alpha_j \alpha_k \int_0^\tau \mathrm{d}s \int_0^s \mathrm{d}u \,  \mathrm{Tr}(\overline{\varrho}[p(s),p(u)])\sin(s\nu_j)\cos(u \nu_k) \Big) \\
&+\mu_0^2 \mathrm{Tr}\big (\overline{\varrho} \,p(s)\, p(u)\big) \mathrm{e}^{i (s-u)\nu_j} + c.c.
\end{align}
The term in the second line is due to the bosonic commutation relations of the quantized field and is negligible with respect to the other one for intense light pulses $|\alpha| \gg 1$. The quantity in the trace can be easily evaluated
\begin{equation}
\mathrm{Tr}\big (\overline{\varrho} \,p(s)\, p(u)\big) = \sum_{\ell m} p_\ell^0 \, \mathrm{e}^{-i (E_m^1-E_\ell^0)(s-u)} \Big| \langle \ell ,0| p |m, 1\rangle \Big|^2 .
\end{equation}
As a result, the time dependence is given by the following integral
\begin{align}
&8i\int_0^\tau \mathrm{d}s \int_0^s \mathrm{d}u \, (-2i) \sin\big((E_m^1-E_\ell^0)(s-u)\big) \sin(s\nu_j)\cos(u \nu_k) = \nonumber \\
&=4\left( \frac{1}{\nu_k -\Delta_{m\ell}} - \frac{1}{\nu_k + \Delta_{m\ell} } \right) \times \nonumber\\
&\times 2 \left( \frac{\sin^2\big(\tau \frac{\nu_j + \nu_k}{2}\big)}{\nu_j + \nu_k}  + \frac{\sin^2\big(\tau \frac{\nu_j - \nu_k}{2} \big)}{\nu_j - \nu_k} - \frac{\sin^2\big(\tau \frac{\nu_j + \Delta_{m\ell}}{2}\big)}{\nu_j + \Delta_{m\ell}} - \frac{\sin^2\big(\tau \frac{\nu_j - \Delta_{m\ell}}{2}\big)}{\nu_j - \Delta_{m\ell}} \right)
\end{align}
with $\Delta_{m\ell}= E_m^1- E_\ell^0$.
By defining the function $D_\tau(x)= \frac{4 \sin^2(\tau x/2)}{x^2} $ we see that the function $\delta_\tau (x) = \frac{1}{2\pi\tau}D_\tau(x) $ is a representation of the Dirac delta in the limit $\tau \to \infty$. Therefore, one finds for the rate of change in transmissivity at frequency $\nu_j$
\begin{equation}
\lim_{\tau \to \infty}\frac{\Gamma^{(2)}(\tau)}{\tau} = -4\pi \mu_0^2 \sum_{\ell m} p_\ell^0 \Big| \langle \ell ,0| p |m, 1\rangle \Big|^2 \delta (\nu_j - \Delta_{m\ell})
\end{equation}
This corresponds to the rate computed through the Fermi golden rule with a minus sign.

\noindent We now perform the same calculation for a state $\overline{\varrho}$ that is of the form
\begin{equation}
\overline{\varrho} = \sum_\ell p^0_\ell \, D | \ell ,0 \rangle \langle \ell, 0 | D^\dag
\end{equation}
for some displacement operator $D= \mathrm{e}^{B b^\dag - B b}$.
This is done in order to take into account the excitation of the vibrational degree of freedom in the sample due to the infrared pump pulse. Indeed, in the following, we model the dynamics induced by the pump as a time-dependent displacement operator acting on the vibrational degree of freedom.
This is a realistic scenario when describing, for instance, stimulated Raman scattering \cite{Glerean2019}. In what follows, all the details about the interaction with the pump are implicitly encoded in the parameter $B_t$, namely, we would not rely on a specific model to predict the functional form of $B_t$

\noindent The calculation can be performed exactly in the same way as before, but the correlation function $\mathrm{Tr}\big (\overline{\varrho} \,p(s)\, p(u)\big)$ reads now
\begin{align}
\mathrm{Tr}\big (\overline{\varrho} \,p(s)\, p(u)\big) &= \sum_{\ell m n v} p_\ell^0 \langle \ell ,0| D^\dag |m, 0\rangle \langle m ,0| p |n, 1\rangle \langle n ,1| p |v, 0\rangle \langle v ,0| D |\ell, 0\rangle \times \nonumber \\
& \times \mathrm{e}^{-i s \Delta_{nm}} \,\mathrm{e}^{i u \Delta_{nv}}.
\end{align}
Integrating in time and performing the same limit as before for the rate one finds 
%\\ \textcolor{red}{details missing here, comment more on this}
\begin{align*}
\lim_{\tau \to \infty} \frac{\Gamma^{(2)}(\tau)}{\tau} & = \mu_0^2 \delta(\nu_j - \overline{\epsilon}- \omega \ell) \sum_n \langle n | D \varrho_\beta D^\dag (b+b^\dag) \mathrm{e}^{-\frac{M}{\omega}(b^\dag -b)} | n+\ell \rangle \times \nonumber \\
& \times \langle n+\ell | \mathrm{e}^{\frac{M}{\omega}(b^\dag -b)}(b+b^\dag) | n \rangle + c.c.
\end{align*}
Therefore, the spectral line $\Gamma_\ell$ reads
\begin{equation}
\Gamma_\ell = \mu_0^2 \sum_n \langle n | D \varrho_\beta D^\dag (b+b^\dag) \mathrm{e}^{-\frac{M}{\omega}(b^\dag -b)} | n+\ell \rangle \langle n+\ell | \mathrm{e}^{\frac{M}{\omega}(b^\dag -b)}(b+b^\dag) | n \rangle .
\end{equation}
We can first discuss the total absorption as in the equilibrium case. The calculation is quite straightforward and the result is
\begin{equation}
\Gamma = \sum_{\ell=-\infty}^{+\infty} \Gamma_\ell = \mu_0^2 \left( \coth\Big(\frac{\beta\omega}{2}\Big) + 4 B_t^2 \right).
\end{equation}
According to this model, the correction to the total absorption is therefore always positive. However, it is interesting to see whether the spectral weight can be shifted or not depending on $B_t$. This is done in the following, computing each single $\Gamma_\ell$.

\noindent Let us start with the case $\ell >0$.
The first matrix element can be conveninetly rewritten exploiting the bosonic commutation relations
\begin{align}
&\langle n | D \varrho_\beta D^\dag (b+b^\dag) \mathrm{e}^{-\frac{M}{\omega}(b^\dag -b)} | n+\ell \rangle = \nonumber \\
=& \mu_0^2\frac{\mathrm{e}^{-\beta \omega n}}{Z_\beta} \langle n | \mathrm{e}^{B_t(\mathrm{e}^{\beta \omega}  b^\dag- \mathrm{e}^{-\beta \omega}b)} (b^\dag + b + 2B_t) \mathrm{e}^{-(\frac{M}{\omega}+B_t)(b^\dag -b)} | n+\ell \rangle = \nonumber \\
=& \mu_0^2\frac{\mathrm{e}^{-\beta \omega n}}{Z_\beta} \langle n | (b^\dag + b -2B_t \cosh(\beta\omega)+ 2B_t) \mathrm{e}^{B_t(\mathrm{e}^{\beta \omega}  b^\dag- \mathrm{e}^{-\beta \omega}b)} \mathrm{e}^{-(\frac{M}{\omega}+B_t)(b^\dag -b)} | n+\ell \rangle = \nonumber \\
=& \mu_0^2\frac{\mathrm{e}^{-\beta \omega n}}{Z_\beta} \mathrm{e}^{-B_t(\frac{M}{\omega}+B_t)\sinh(\beta\omega)} \langle n | \Big( b+b^\dag-4B_t \sinh^2\Big(\frac{\beta\omega}{2}\Big) \Big) \mathrm{e}^{b^\dag x + b y} | n+\ell \rangle
\end{align}
where the coefficients $x$ and $y$ are defined as follows
\begin{equation}\label{xy}
x= -\frac{M}{\omega} + B_t (\mathrm{e}^{\beta\omega}-1), \quad y= \frac{M}{\omega} + B_t (1-\mathrm{e}^{-\beta\omega}).
\end{equation}
Finally one has
\begin{align}
&\langle n | D \varrho_\beta D^\dag (b+b^\dag) \mathrm{e}^{-\frac{M}{\omega}(b^\dag -b)} | n+\ell \rangle = \nonumber \\
=& \mu_0^2\frac{\mathrm{e}^{-\beta \omega n}}{Z_\beta} \mathrm{e}^{-\frac{M^2}{2\omega^2}(1+2\frac{\omega B_t}{M}(1+\frac{\omega B_t}{M})(1-\mathrm{e}^{-\beta \omega}))} \langle n | \Big( b+b^\dag-4B_t \sinh^2\Big(\frac{\beta\omega}{2}\Big) \Big) \mathrm{e}^{x b^\dag}\mathrm{e}^{y b} | n+\ell \rangle
\end{align}

Using again the properties of bosonic operators one can write for $m > n$
\begin{equation}
\langle n | \mathrm{e}^{x b^\dag} \mathrm{e}^{y b}   | m \rangle = \frac{\sqrt{n!}}{\sqrt{m!}} y^{m-n} L_{n}^{m-n} (-xy),
\end{equation}
where the product $-xy$ explicitly reads
\begin{equation}
-xy = \frac{M^2}{\omega^2} - 4 B_t \Big( \frac{M}{\omega} + B_t \Big)  \sinh^2\Big( \frac{\beta \omega}{2} \Big).
\end{equation}
The matrix element $\langle n | \Big( b+b^\dag - 4 B_t \sinh^2\big(\frac{\beta \omega}{2}\big) \Big) \mathrm{e}^{x b^\dag} \mathrm{e}^{y b}  | n+\ell \rangle$ then reads
\begin{equation}
\sqrt{n} \langle n-1 | X | n+ \ell \rangle + \sqrt{n+1} \langle n+1 | X | n+ \ell \rangle -4B_t \sinh^2\Big(\frac{\beta\omega}{2} \Big) \langle n | X | n+ \ell \rangle,
\end{equation}
where $X= \mathrm{e}^{x b^\dag} \mathrm{e}^{y b} $, and can be rewritten accordingly as
\begin{equation}
\frac{\sqrt{n!}}{\sqrt{(n+\ell)!}} \Big[ y^{\ell-1} \Big( y^2 L_{n-1}^{\ell+1}(-xy) + (n+1)L_{n+1}^{\ell-1}(-xy) \Big)  -4B_t \sinh^2\Big(\frac{\beta\omega}{2} \Big) y^\ell L_{n}^{\ell}(-xy)  \Big].
\end{equation}
Using the recurrence relations for the Laguerre polynomials one finds
\begin{equation}
\frac{\sqrt{n!}}{\sqrt{(n+\ell)!}} y^{\ell-1} \Big[ \Big(  (\ell +xy)  -4B_t y \sinh^2\Big(\frac{\beta\omega}{2} \Big) \Big) L_{n}^{\ell}(-xy)  +  (y^2 + xy) L_{n-1}^{\ell +1}(-xy) \Big].
\end{equation}
The first term can be summed as in the time-independent case. In particular, one finds
\begin{equation}
\sum_{n=0}^{\infty}\frac{n!}{(n+\ell)!} \mathrm{e}^{-\beta \omega n} L_{n}^{\ell}\Big(-xy\Big)L_{n}^{\ell} \Big(\frac{M^2}{\omega^2}\Big) = \frac{\mathrm{e}^{-(\frac{M^2}{\omega^2}-xy)\frac{1}{\mathrm{e}^{\beta\omega}-1}}}{(1-\mathrm{e}^{-\beta\omega})\mathrm{e}^{-\frac{\beta\omega\ell}{2}}\Big(-xy\frac{M^2}{\omega^2} \Big)^{\frac{\ell}{2}}} I_\ell \Bigg( \frac{\big(-xy\frac{M^2}{\omega^2} \big)^{\frac{1}{2}}}{\sinh(\beta\omega/2)} \Bigg)
\end{equation}
The other one can be also treated explicitly using the property 
\begin{equation}
L_{n-1}^{\ell+1}(z)= -\frac{\mathrm{d}}{\mathrm{d}z} L_{n}^{\ell}(z).
\end{equation}
Indeed one can write 
\begin{align}
&\sum_{n=0}^{\infty}\frac{n!}{(n+\ell)!} t^n L_{n-1}^{\ell+1}(z)L_{n}^{\ell}(w) = \nonumber \\
& =-\frac{\mathrm{d}}{\mathrm{d}z} \sum_{n=0}^{\infty}\frac{n!}{(n+\ell)!} t^n L_{n}^{\ell}(z)L_{n}^{\ell}(w) = \nonumber \\
&= \frac{(zwt)^{-\ell/2}}{1-t} \mathrm{e}^{-\frac{(z+w)t}{1-t}} \Big[ \frac{t}{1-t} \,I_\ell \left(\frac{2\sqrt{zwt}}{1-t}\right) - \frac{\sqrt{wt}}{\sqrt{z}(1-t)} \,I_{\ell+1} \left(\frac{2\sqrt{zwt}}{1-t}\right) \Big].
\end{align}
The exchange of derivative and summation is allowed by the uniform convergence of the series in compact sets $|z|<a, |w|<b$. In order to prove uniform converegence it is sufficient to notice that
\begin{equation}
|L_n^\ell (z)| \leq L_n^\ell(-a) , \quad |z|\leq a.
\end{equation}
Indeed, the quantity $|a_n|$ is bounded as follows
\begin{equation}
|a_n|= \frac{n!}{(n+\ell)!} t^n |L_n^\ell(z)||L_n^\ell(w)| \leq \frac{n!}{(n+\ell)!} t^n L_n^\ell(-a) L_n^\ell(-b) \equiv M_n
\end{equation}
and $\sum_n M_n < \infty $. This proves the uniform convergence according to the Weierstrass criterion.
The sum reads explicitly
\begin{align}
&\sum_{n=0}^{\infty} \frac{n!}{(n+\ell)!} t^n L_{n-1}^{\ell+1}\Big(-xy\Big)L_{n}^{\ell} \Big(\frac{M^2}{\omega^2}\Big) = \nonumber \\
=& \frac{\mathrm{e}^{-(\frac{M^2}{\omega^2}-xy)\frac{1}{\mathrm{e}^{\beta\omega}-1}}}{(1-\mathrm{e}^{-\beta\omega})\mathrm{e}^{-\frac{\beta\omega\ell}{2}}\Big(-xy\frac{M^2}{\omega^2} \Big)^{\frac{\ell}{2}}} \left[ \frac{1}{\mathrm{e}^{\beta\omega}-1} I_\ell \Bigg( \frac{\big(-xy\frac{M^2}{\omega^2} \big)^{\frac{1}{2}}}{\sinh(\beta\omega/2)} \Bigg) - \frac{\frac{M}{\omega}}{2\sinh(\beta\omega/2)(-xy)^{1/2}} I_{\ell+1} \Bigg( \frac{\big(-xy\frac{M^2}{\omega^2} \big)^{\frac{1}{2}}}{\sinh(\beta\omega/2)} \Bigg) \right]
\end{align}

A similar calculation can be performed in the case $\ell \leq 0$ and, in the end, each single $\Gamma_\ell$ reads
\begin{align}
\Gamma_\ell &= \mu_0^2 (\ell -A) \frac{ \mathrm{e}^{-A \coth(C)} \mathrm{e}^{C\ell}}{\Big(1-4B(1+B)\sinh^2(C) \Big)^{\ell/2}}   \Big( 1+ B(1-\mathrm{e}^{-2C})  \Big)^{\ell-1} A^{-1}  \times \nonumber \\
&\times \Bigg[ \Big( \ell -A +AB\big(1+\mathrm{e}^{-2C} + 2B(1-\mathrm{e}^{-2C})\big) \Big) I_\ell\Big(\frac{A (1-4B(1+B)\sinh^2(C) )^{1/2}}{\sinh(C)} \Big) + \nonumber \\
&- AB\frac{\sinh(2C)(1+B(1-\mathrm{e}^{-2C}))}{\sinh(C)(1-4B(1+B)\sinh^2(C) )^{1/2}} I_{\ell+1}\Big(\frac{A (1-4B(1+B)\sinh^2(C) )^{1/2}}{\sinh(C)} \Big) \Bigg], \quad \ell>0,
\label{eq_final}
\end{align}
\begin{align}
\Gamma_\ell &= \mu_0^2 (|\ell| + A) \frac{ \mathrm{e}^{-A \coth(C)} \mathrm{e}^{C\ell}}{\Big(1-4B(1+B)\sinh^2(C) \Big)^{|\ell|/2}}   \Big( 1- B(\mathrm{e}^{2C}-1)  \Big)^{|\ell|-1} A^{-1}  \times \nonumber \\
&\times \Bigg[ \Big( |\ell| +A -2AB\big(\frac{\mathrm{e}^{2C}\sinh(2C)}{\mathrm{e}^{2C}-1} - B(\mathrm{e}^{2C}-1)\big) \Big) I_{|\ell|}\Big(\frac{A (1-4B(1+B)\sinh^2(C) )^{1/2}}{\sinh(C)} \Big) + \nonumber \\
&- AB\frac{-\sinh(2C) + 2B \sinh^2(C)(\mathrm{e}^{2C}+1)}{\sinh(C)(1-4B(1+B)\sinh^2(C) )^{1/2}} I_{|\ell|+1}\Big(\frac{A (1-4B(1+B)\sinh^2(C) )^{1/2}}{\sinh(C)} \Big) \Bigg], \quad \ell\leq 0,
\label{eq_final2}
\end{align}
where we defined the three adimensional parameters $A,B,C$ as
\begin{equation}
A= \frac{M^2}{\omega^2}, \quad B= B_t \frac{\omega}{M},  \quad C= \frac{\beta \omega}{2}.
\label{eq_parameters}
\end{equation}
These three parameters completely specify the model in the approximation we used. The parameter $A$ quantifies the phonon-electron coupling and therefore the displacement of the nuclear positions in the electronic excited state. The parameter $B$ instead is the dynamical one, related to the pump pulse, and gives the ratio between the light-induced instantaneous displacement and the equilibrium one due to the electron-phonon coupling. The third parameter, $C$, specifies the temperature in units of the phonon frequency.
In Supplementary Figure \ref{ABC_dependency}.a-c distributions of the $\Gamma_\ell$ are plotted for different values of the parameters $A$, $B$, and $C$ keeping, for each subfigure, two parameters constant, in order to show the trend for any variation of the third one. In particular, we have assumed that: (i) the relevant phonon mode has a frequency $\omega=16$ meV, (ii) the coupling constant $M=\omega$ in order to match the width of the experimental outcome and (iii) $T=300$ K ($C=0.31$) for Supplementary Fig.\ref{ABC_dependency}.a-b. One can also numerically compute the central energy ($<E>$) and the energy bandwidth ($\sigma_E$) of these distributions. Thus, Figure \ref{ABC_dependency}.a shows that a stronger e-p couplings increases the number of possible transitions, which broadens the distribution, and pushes it toward higher energy. Figure \ref{ABC_dependency}.b shows that a higher $B_t$ mostly displaces the overall distribution toward higher energy. Figure \ref{ABC_dependency}.c shows that a lower temperature mostly narrows the distribution.

\begin{figure}[h!]
\centering
\includegraphics[scale=0.28]{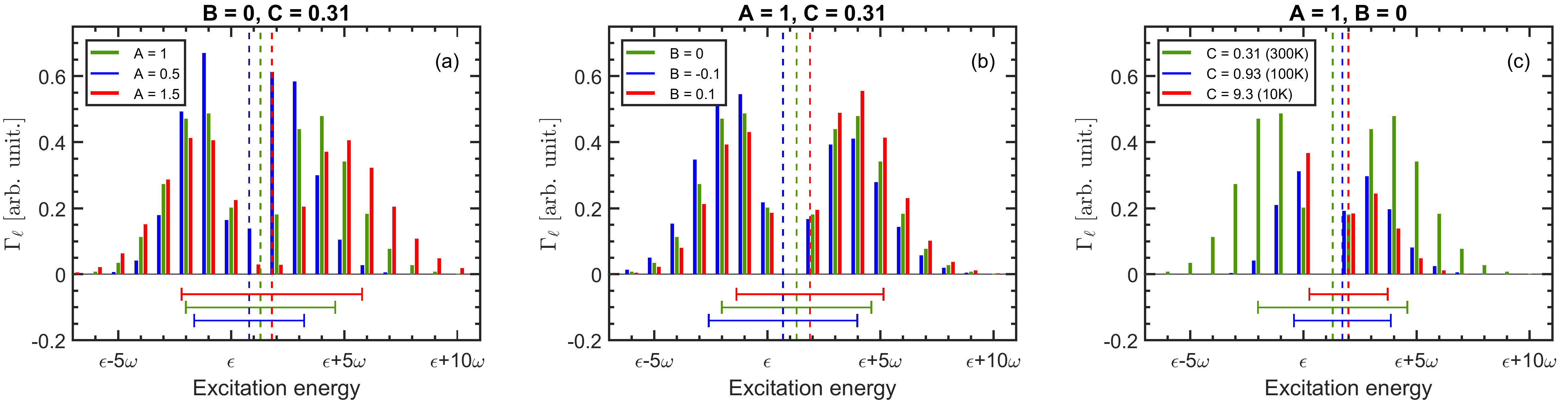}
\caption{Absorption probability distributions (eq.\ref{eq_final}-\ref{eq_final2}) for a variation of the three parameters (eq.\ref{eq_parameters}): (a) A (strength of the e-p couplings), (b) B (amplitude of the displacement) and (c) C (temperature). For each graph, the values of the two other parameters are kept constant and indicated on the top of the graph. Moreover the central energy (resp. the energy bandwidth) is represented by a vertical dashed line (resp. a horizontal line below the graph).}
\label{ABC_dependency}
\end{figure}

\noindent Then, we can also study the trends of the central energy and the energy bandwidth of the distribution as a function of a wide range of temperatures and displacements, in the case $M=\omega=16$ meV. The results are plotted in Supplementary Fig.\ref{T_B_dependencies_large_scale}.a-d. Each point of these curves can be seen as if the sample was in a different initial condition $T_{eq}$ and $B_{t,eq}$. The results confirm the overall trends which have been mentioned above. Indeed, the observed trends are mostly monotonic and most of them do not change of behavior (increase or decrease) with a change of the secondary parameter: $B_t$ for the T-dependency (subfigures a-b) and $T$ for the $B_t$-dependency (subfigures c-d). Interestingly, the central energy $<E>$ as a function of the temperature displays a different behavior depending on $B_t$: $<E>$ decreases with $T$ if $B_t \lesssim 0.25$ but $<E>$ increases with $T$ if $B_t \gtrsim 0.35$. As matter of fact, we know that the central energy of the d-d transitions should decrease when increasing the sample temperature. Thus, it gives us a range of validity of the model and, in the following, we choose $B_{t,eq}=0.1$ to respect this experimental evidence.

\begin{figure}[h!]
\centering
\includegraphics[scale=0.35]{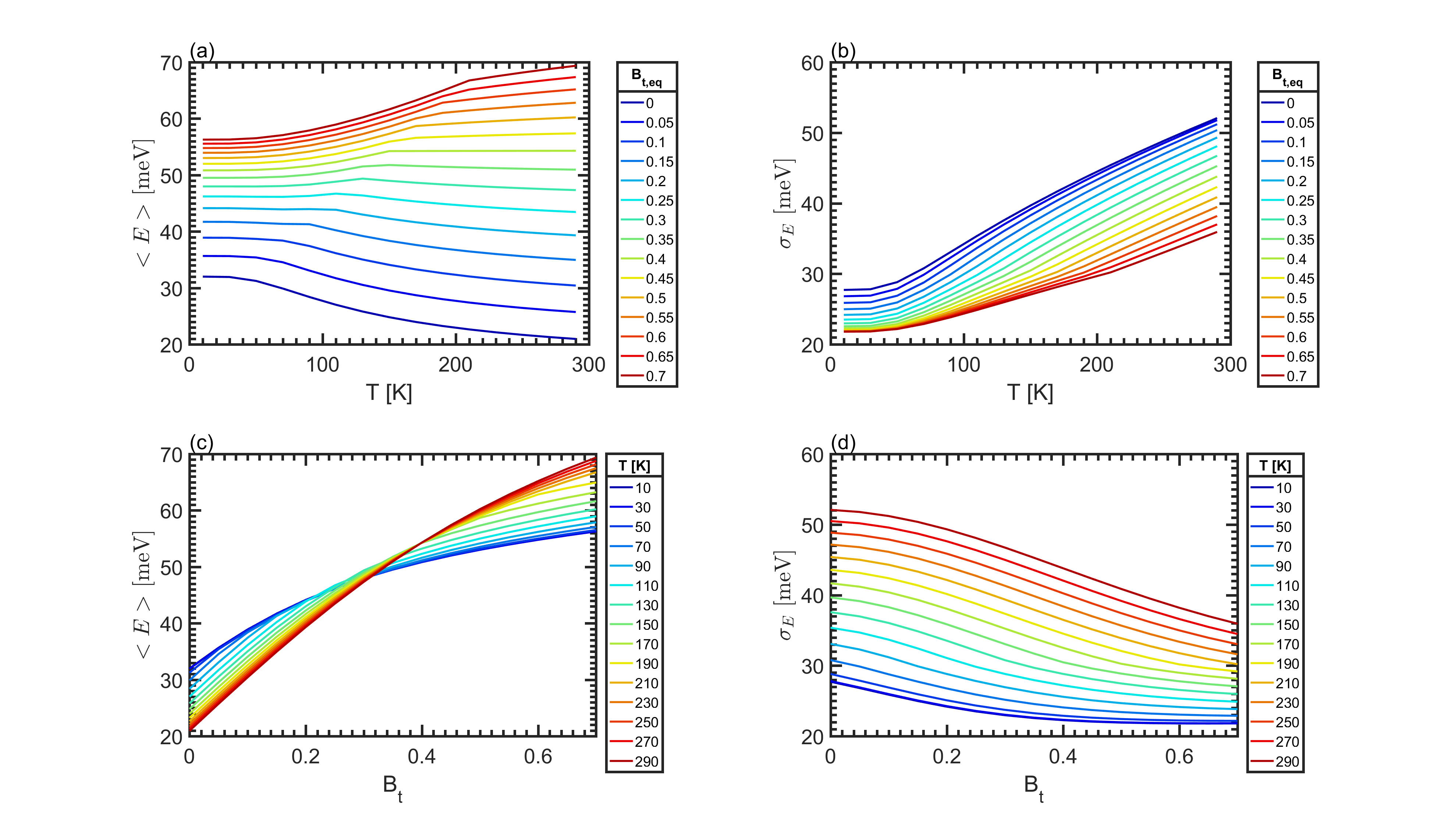}
\caption{Central energy as a function of (a) the temperature and (c) the displacement. Energy bandwidth as a function of (b) the temperature and (d) the displacement. In each graph, a given color corresponds to a given secondary parameter: $B_t$ for a-b and $T$ for c-d.}
\label{T_B_dependencies_large_scale}
\end{figure}

\begin{figure}[h!]
\centering
\includegraphics[scale=0.35]{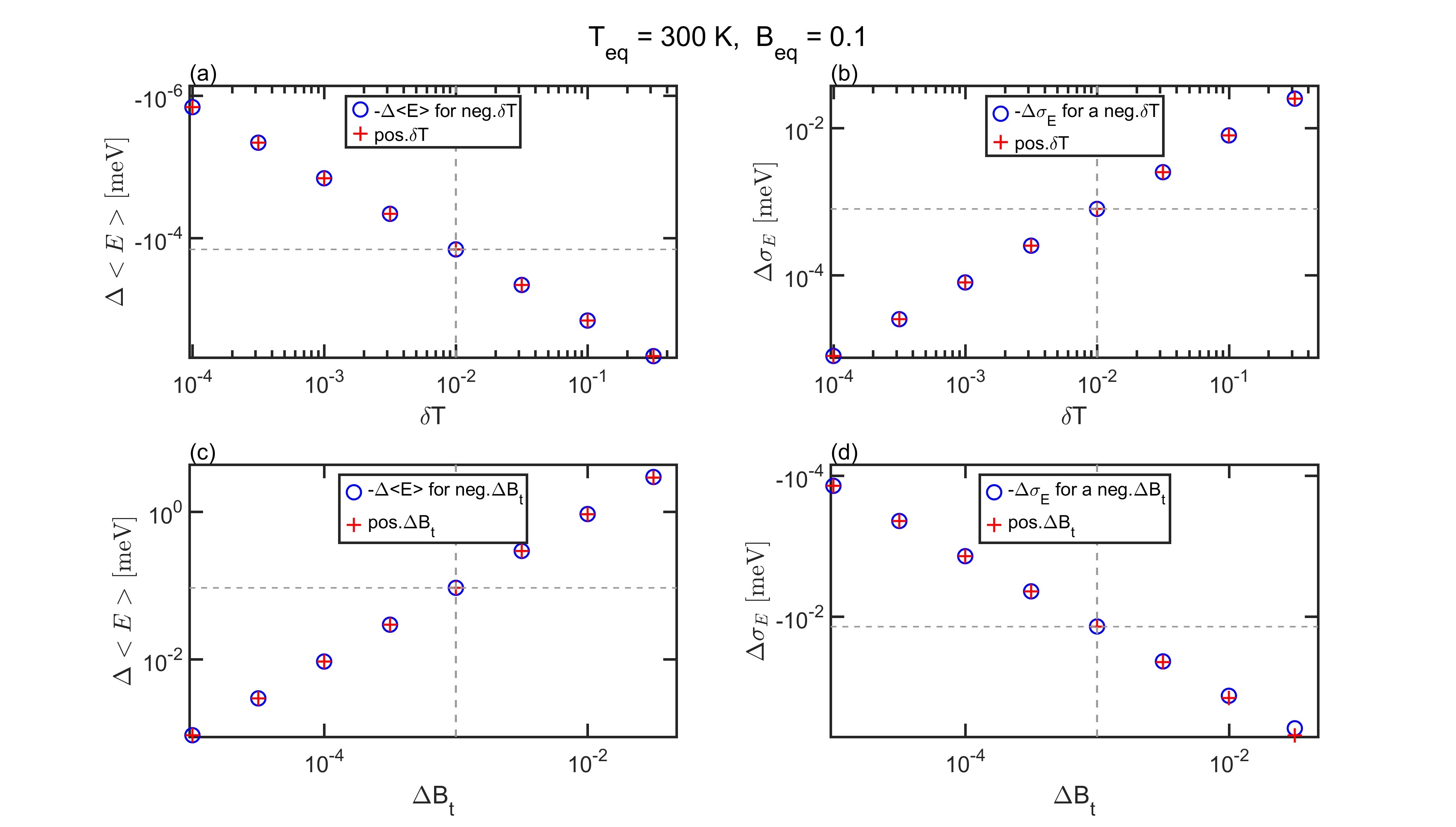}
\caption{Central energy shift as a function of (a) the temperature and (c) the displacement. Energy bandwidth shift as a function of (b) the temperature and (d) the displacement. Positive parameter variations are represented by red cross and in the case of a negative parameter variation, the opposite of the computed shift is plotted as a function of the absolute value of the variation (blue circle).}
\label{T_B_dependcies_300K_01}
\end{figure}

\noindent As explained in the main text, we are particularly interested in comparing the impact of a temperature variation around an equilibrium value $T_{eq}$ with respect to the effects of a displacement variation around an equilibrium $B_{t,eq}$. Thus, we have studied the trend of the central energy and the energy bandwidth as a function of different variations of the temperature (Supplementary Fig.\ref{T_B_dependcies_300K_01}.a-b) or of the displacement (Supplementary Fig.\ref{T_B_dependcies_300K_01}.c-d)) around some experimentally relevant parameters: $M=\omega=16$ meV, $T_{eq}=300$ K, $B_{t,eq}=0.1$. As it is shown by all the figures, the trend is rather linear for positive or negative variations of the parameters $\delta T$ and $\Delta B_t$. The logarithm scale allows to better estimate $\Delta <E>$ or $\Delta \sigma_E$ as a function of $\delta T$ or $\Delta B_t$. In particular, we can observe that the retrieved $\Delta <E>$ or $\Delta \sigma_E$ for $\delta T =10^{-2}$ K is one to two order of magnitude lower than the retrieved variation for $\Delta B_t =10^{-3}$, which would be equivalent to a pump-induced displacement of $10^{-4} \si{\angstrom}$. This again demonstrates that the displacement variation around an equilibrium condition has a major role to play in the subsequent dynamical variations of the d-d transitions electronic properties. Note that the computation of these variations at $T_{eq}=8$ K confirms this trend (taking into account $\delta T =1$ K and $\Delta B_t =10^{-3}$).

\subsection{Molecular electronic excitations}
\label{molecule}

In our model we considered an interaction Hamiltonian between the light and the sample that is proportional to the phonon position operator $b+b^\dag$. In order to justify this choice we recall here some basic notions of molecular physics.

\noindent Consider the wave function of a molecule in the Born-Oppenheimer approximation
\begin{equation}
\Phi_{\ell \nu} (r, R) = \psi_{\ell}(r;R)\chi_{\nu, \ell}(R),
\end{equation}
where the electronic wave function $\psi_{\ell}(r;R)$ is a solution of the clamped-nuclei Schr\"odinger equation and depends parametrically of $R$, and the nuclear wave function $\chi_{\nu, \ell}(R)$ is the $\nu-th$ level eigenfunction relative to the harmonic potential corresponding to level $\ell$.

\noindent The transition probability rate between two electronic states, labelled by $\ell=0$ and $\ell=1$, and vibrational levels $n \to m$, can be computed according to the Fermi Golden rule and reads
\begin{equation}
\Gamma_{0 \to 1}(\omega)= |\langle \nu=m,\ell=1| \mu |\nu=n, \ell=0\rangle |^2 \delta(\omega - E_{m-n})
\end{equation}
where the energy of the transition is $E_{m-n}=\epsilon + \omega_0(m-n) $ and the operator $\mu= e Z\cdot R- e r= \mu_{el}(r) + \mu_{nu}(R)$ is the electric dipole moment containing both nuclear and electronic coordinates. The matrix element reads
\begin{align}\label{transmat}
&\langle \nu=m,\ell=1| \mu |\nu=n, \ell=0\rangle = \int \mathrm{d}R \, \mathrm{d}r \,\psi^{*}_1 (r;R)\chi^{*}_{m,1}(R)(\mu)\psi_{0}(r;R)\chi_{n,0}(R) \nonumber \\
&= \int \mathrm{d}R \chi^{*}_{m,1}(R) \, \mu_{nu}(R) \,\chi_{n,0}(R) \, \int \mathrm{d}r \psi^{*}_1 (r;R)  \psi_{0}(r;R) + \nonumber\\
&\quad + \int \mathrm{d}R \chi^{*}_{m,1}(R) \, \chi_{n,0}(R) \, \int \mathrm{d}r \psi^{*}_1 (r;R) \mu_{el}(r) \psi_{0}(r;R).
\end{align}
The first term is vanishing because the electronic wave functions form an orthonormal basis in the electronic Hilbert space for each fixed position of the nuclei $R$. For the same reason, we can expand a generic $\psi_{n}(r; R)$ in terms of the wave functions at the equilibrium position $\psi_{m}(r; R_0)$
\begin{equation}
\psi_{n}(r; R) = \sum_{m} c_{nm}(R-R_0) \psi_{m}(r; R_0).
\end{equation}
For later convenience, we can define the integral 
\begin{equation}
I_{mn} := \int \mathrm{d}r \psi^{*}_m (r;R_0) \mu_{el}(r) \psi_n (r;R_0).
\end{equation}
Using this notation, the integral over the electronic coordinates in the second term of \eqref{transmat} becomes
\begin{align}\label{elint}
\int \mathrm{d}r \psi^{*}_1 (r;R) \mu_{el}(r) \psi_{0}(r;R) &= c^{*}_{11} c_{00} I_{10} + \sum_{m \neq 1} c^{*}_{1m} c_{00} I_{m0} + \nonumber \\
&+ \sum_{n \neq 0} c^{*}_{11} c_{0n} I_{1n} + \sum_{m \neq 1, n\neq 0} c^{*}_{1m} c_{0n} I_{mn}.
\end{align}
If the bare electronic transition is allowed by symmetry selection rules, then the first term $I_{10}$ is the dominant contribution to the matrix element. On the contrary, if symmetry forbids the transition in equilibrium, namely $I_{10}=0$, the other contributions become relevant \cite{Bransden2003, Demtroder2006}. In particular, for small displacement from the equilibrium position we can expand the coefficients $c$ in a Taylor series
\begin{align*}
&c_{nn} = 1+ \alpha_n (R-R_0) + O((R-R_0)^2), \\
&c_{nm}= \alpha_{nm} (R-R_0) + O((R-R_0)^2), \quad m \neq n
\end{align*}
with some complex parameters $\alpha$. The dominant terms in \eqref{elint} are then
\begin{equation}
\int \mathrm{d}r \psi^{*}_1 (r;R) \mu_{el}(r) \psi_{0}(r;R) \simeq (R-R_0)\left( \sum_{m \neq 1} \alpha^{*}_{1m} I_{m0} + \sum_{n \neq 0} \alpha_{0n} I_{1n} \right).
\end{equation}
Coming back to the full expression for the matrix element \eqref{transmat} we get
\begin{equation}
\langle \nu=m,l=1| \mu |\nu=n, l=0\rangle = \left( \sum_{m \neq 1} \alpha^{*}_{1m} I_{m0} + \sum_{n \neq 0} \alpha_{0n} I_{1n} \right) \times \int \mathrm{d}R \chi^{*}_{m,1}(R) (R-R_0) \chi_{n,0}(R) .
\end{equation}
Therefore, our model is consistent with the previous findings, because the height of the spectral lines is proportional to the modulus squared of the following matrix element 
\begin{equation}
\langle m | \mathrm{e}^{\frac{M}{\omega} (b^\dag - b)} (b+b^{\dag}) | n \rangle .
\end{equation}

\subsection{Estimate of the relevant length scale}
\noindent
In order to estimate the relevant length scale in our model, we consider a displaced harmonic potential of the form
\begin{equation}
V = \frac{1}{2}m\omega^2 x^2 + \lambda x,
\end{equation}

\noindent where $\omega$ is the frequency, $m$ is the mass of the oscillator and $\lambda$ represents the coupling. In natural units ($c=\hbar=k_b=1$), mass and frequency have the same physical units of energy, say $eV$, while the position $x$ is expressed in $eV^{-1}$. For dimensional consistency $\lambda$ has units of $eV^2$. The minimum of the harmonic trap $V_{min}$ and the corresponding position $x_{min}$ are

\begin{equation}
x_{min}= -\frac{\lambda}{m \omega^2}, \quad V_{min}= -\frac{1}{2}\frac{\lambda^2}{m \omega^2}.
\end{equation}
By comparing the energy shift with the calculation done in the manuscript we can infer the relation between $\lambda$ and $M$

\begin{equation}
\lambda = \sqrt{2m\omega} M ,
\end{equation} 

\noindent and rewrite the position of the minimum in terms of $M$

\begin{equation}
x_{min}= - \frac{M}{\omega} \frac{\sqrt{2}}{\sqrt{m\omega}}.
\end{equation}

\noindent This is also consistent with the evaluation of the average position in the displaced ground state computed using the quantum model 

\begin{equation}
x_{ground} = \frac{1}{\sqrt{2m \omega}} \langle b + b^\dag \rangle = -2 \frac{M}{\omega} \frac{1}{\sqrt{2m \omega}} = - \frac{M}{\omega} \frac{\sqrt{2}}{\sqrt{m\omega}}.
\end{equation}

\noindent In order to get numerical values for the quantity $x_{min}$ we assume $\omega= 16 meV$ and consider the mass of Copper as reference  $m= 63.546\, a.m.u.= 5.9 \cdot 10^{10} eV$.

\noindent It turns out that 

\begin{equation}
\frac{\sqrt{2}}{\sqrt{m \omega}} \simeq 0.46 \cdot 10^{-4} eV^{-1}
\end{equation}

\noindent and using the conversion factor $1 eV^{-1}= 1.97 \cdot 10^{-7} m$ we can find the connection between the physical displacement $\Delta x$ and the adimensional parameter $\frac{M}{\omega}$

\begin{equation}
\Delta x \sim \frac{M}{\omega} \cdot 0.9 \cdot 10^{-11} \,m =  \frac{M}{\omega}  0.09  \si{\angstrom}.
\end{equation}

\noindent The role of $B_t$ is analogous to that of $M/\omega$ in the previous equation.

\subsection{Anharmonic coupling}
\label{anharmonic_coupling}
We stated in section \ref{longtime} that oscillations in the transmissivity map are observed at long timescale at the frequency of the low energy Raman active $A_g$ mode (187 cm$^{-1}$) \cite{Popovic1995}. However, given the pump frequency (around 9 $\mu$m) and polarization (along the c axis), the mostly coupled phonon mode should be the high frequency IR active mode $B_{2u}$ (720 cm$^{-1}$). The observed phenomenology can be explained by means of anharmonic coupling between the high frequency IR active mode and the low frequency Raman active mode as given by the following Hamiltonian \cite{Forst2011} : 
\begin{equation}
H_A = - N A Q^2_{IR} Q_{RS},
\end{equation}
where $A$ is some anharmonic coupling constant, $N$ is the number of cells in the lattice, $Q_{IR}$ is the coordinate of the IR active mode and $Q_{RS}$ is the coordinate of the Raman active mode. As a consequence, the equation of motion for the coordinate $Q_{RS}$ reads
\begin{equation}
\frac{\mathrm{d}^2 Q_{RS}}{\mathrm{d}t^2}(t) + \Omega^2_{RS} Q_{RS}(t) = A Q^2_{IR}(t),
\end{equation}
where $\Omega_{RS}$ is the frequency of the Raman mode, so that the fast oscillation of the high energy mode acts as a force field for the low frequency one. In particular, the pump field induces the following evolution of the IR active mode coordinate
\begin{equation}
Q_{IR}(t) = \int_{-\infty}^{\infty}F(\tau)\mathrm{d}\tau  \,\frac{e^* E_0}{\Omega_{IR}\sqrt{M_{IR}}}\cos(\Omega_{IR}t),
\end{equation}
where $F(\tau)$ is the pulse envelope, $E_0$ is the electric field amplitude, $\Omega_{IR}$ is the frequency, $M_{IR}$ is the reduced mass and $e^*$ is the effective charge. Therefore, the coordinate $Q_{RS}$ evolves in time as
\begin{equation}
Q_{RS}(t)= \frac{A}{2\Omega^2_{RS}}\left[ \int_{-\infty}^{\infty}F(\tau)\mathrm{d}\tau  \right]^2 \frac{(e^* E_0)^2}{M_{IR}\Omega^2_{IR}} \big(1-\cos(\Omega_{RS}t)\big).
\end{equation}
Assuming a much longer lifetime for the low frequency phonon mode, we can explain the long-lived oscillations of transmissivity at the frequancy $\Omega_{RS}$.

\section{DFT calculation}
\label{section_DFT_calc}
We support our interpretation of the experimental observations using DFT (Density Functional Theory) simulations of the sample CuGeO$_3$. In particular, after obtaining the optimized structure we compared the ground state properties (band structure, insulating gap, antiferromagnetism of Cu chains) with the existing literature. Finally, we performed the calculation of the phonon modes of the crystal and computed the force field on the octahedron by displacing the atoms along the mode $B_{2u}$.

\subsection{Ground state calculations}

The structure optimization has been performed using the dedicated software QUANTUM ESPRESSO (QE) \cite{Giannozzi_2009}. The lattice parameters have been taken from the literature \cite{Popovic1995} and the initial data for the atomic positions inside the unit cell have been recovered from the Materials Project website https://materialsproject.org/materials/mp-21344.
We used the PBE functional \cite{PBEfunctional} and pseudo-dojo pseudopotentials \cite{pseudodojo} generated with the same functional. 

\begin{figure}[h!]
  \begin{tabular}{ccc}
   \includegraphics[scale=1]{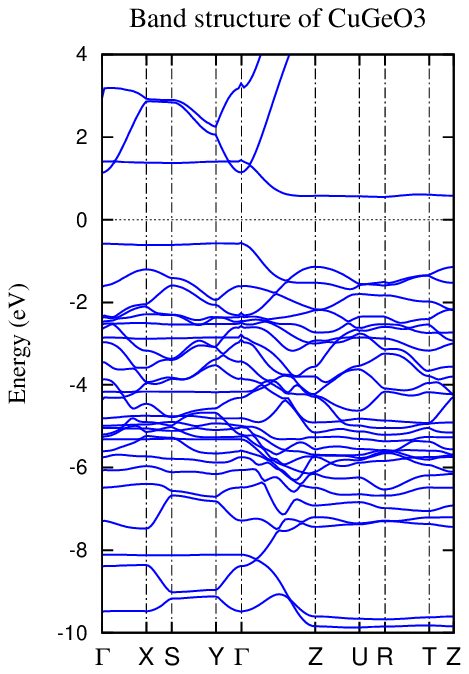}  & \includegraphics[scale=1]{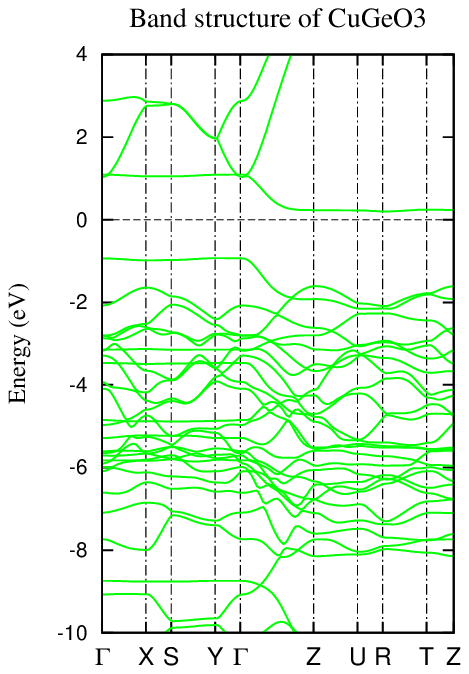} & \includegraphics[scale=0.57]{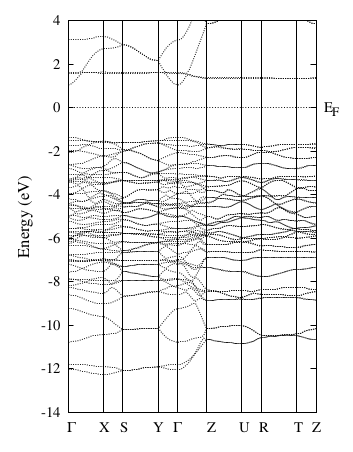}  \\ 
      (a)  & (b) & (c)  \\ 
\end{tabular}
\caption{(a) Band structure of Copper Germanate obtained with \textsf{octopus}, using a spin-polarized PBE+U calculation, with pseudo-dojo pseudopotentials. A U value of $6.7$ eV has been considered in order to compare results with \cite{Wu_1999}. (b) Band structure of Copper Germanate obtained with QE, using a spin-polarized PBE+U calculation, with pseudodojo pseudopotentials. A U value of $6.7$ eV has been considered in order to compare results with \cite{Wu_1999}. (c) Band structure of Copper Germanate obtained in \cite{Wu_1999}, using LSDA+U.}
\label{qe_band}
\label{Wu_band}
\label{oct_band}
\end{figure}

\noindent As a first step, we reproduced the known results for the ground state properties of Copper Germanate taking as a reference \cite{Wu_1999}. In particular, as discussed there, we verified that standard DFT calculations fail to predict the insulating band gap of this material and a more refined treatment is needed. We used a spin-polarized calculation with a Hubbard correction DFTS+U \cite{dftu}. 
We performed the calculation with both QE and \textsf{octopus} \cite{octopus2003, octopus2006}, another DFT dedicated software. The obtained band structures (Supplementary Fig.\ref{oct_band}.a and Supplementary Fig.\ref{qe_band}.b) are compatible and in turn they are in agreement with the result presented in \cite{Wu_1999} that we report here for convenience in Supplementary Fig.\ref{Wu_band}.c.

\noindent Moreover, the spin-polarized calculation correctly converges to a solution with a finite atomic magnetic moment between $0.52\, \mu_B$ (QE) and $0.66\,\mu_B$ (\textsf{octopus}) in the antiferromagnetic Cu chain that is a bit lower than $0.76\,\mu_B$, the value found in \cite{Wu_1999}. However, the different methods that are used there can justify this mismatch.

\subsection{Phonons}
The calculation of the phonon modes has been performed with the software QE. The diagonalization of the dynamical matrix has been performed without the Hubbard correction because this function is currently not available in the software. The result is shown in the Table \ref{phonon_modes}.
\begin{table}[h!]
\begin{center} 
\caption{Phonon modes retrieved with QE.}
\label{phonon_modes}
\begin{tabular}{|c|c|c|}
   \hline
   frequency  & symmetry & IR vs R  \\ 
      \quad   (cm$^{-1}$) &          &  \\ \hline
             -41.9   & $B_{3u}$  &   I \\ \hline
             -22.8   & $B_{2u}$  &   I \\ \hline
              31.9   & $B_{1u}$  &   I \\ \hline
              60.2   & $B_{1u}$  &   I \\ \hline
              79.9   & $A_u$   &  $ $   \\ \hline
             115.4   & $B_{1g}$  &   R \\ \hline
             120.0   & $B_{2g}$  &   R \\ \hline
             124.7   & $B_{3u}$  &   I \\ \hline
             157.6   & $B_{2u}$  &   I \\ \hline
             178.4   & $A_g $  &   R \\ \hline
             191.0   & $B_{1u}$  &   I \\ \hline
             224.0   & $B_{2g}$  &   R \\ \hline
             262.2   & $B_{3u}$  &   I \\ \hline
             272.2   & $B_{1u}$  &   I \\ \hline
             284.8   & $B_{3u}$  &   I \\ \hline
\end{tabular}
\begin{tabular}{|c|c|c|}
 \hline
   frequency  & symmetry & IR vs R  \\ 
      \quad   (cm$^{-1}$) &          &  \\ \hline             
             289.5   & $A_g$   &   R \\ \hline
             341.8   & $B_{1u}$  &   I \\ \hline
             366.3   & $B_{2g}$  &   R \\ \hline
             377.3   & $B_{3g}$  &   R \\ \hline
             395.1   & $B_{1g}$  &   R \\ \hline
             547.7   & $A_u$   &  $ $  \\ \hline
             550.5   & $B_{2u}$  &   I \\ \hline
             582.2   & $A_g$   &   R \\ \hline
             586.6   & $B_{3u}$  &   I \\ \hline
             686.3   & $B_{1u}$  &   I \\ \hline
             700.5   & $B_{1g}$  &   R \\ \hline
             702.6   & $B_{2u}$  &   I \\ \hline
             733.7   & $B_{2g}$  &   R \\ \hline
             745.2   & $B_{3u}$  &   I \\ \hline
             781.6   & $A_g $  &   R \\ \hline
\end{tabular}
\end{center}
\end{table}

\begin{figure}[h!]
    \centering \includegraphics[scale=0.5]{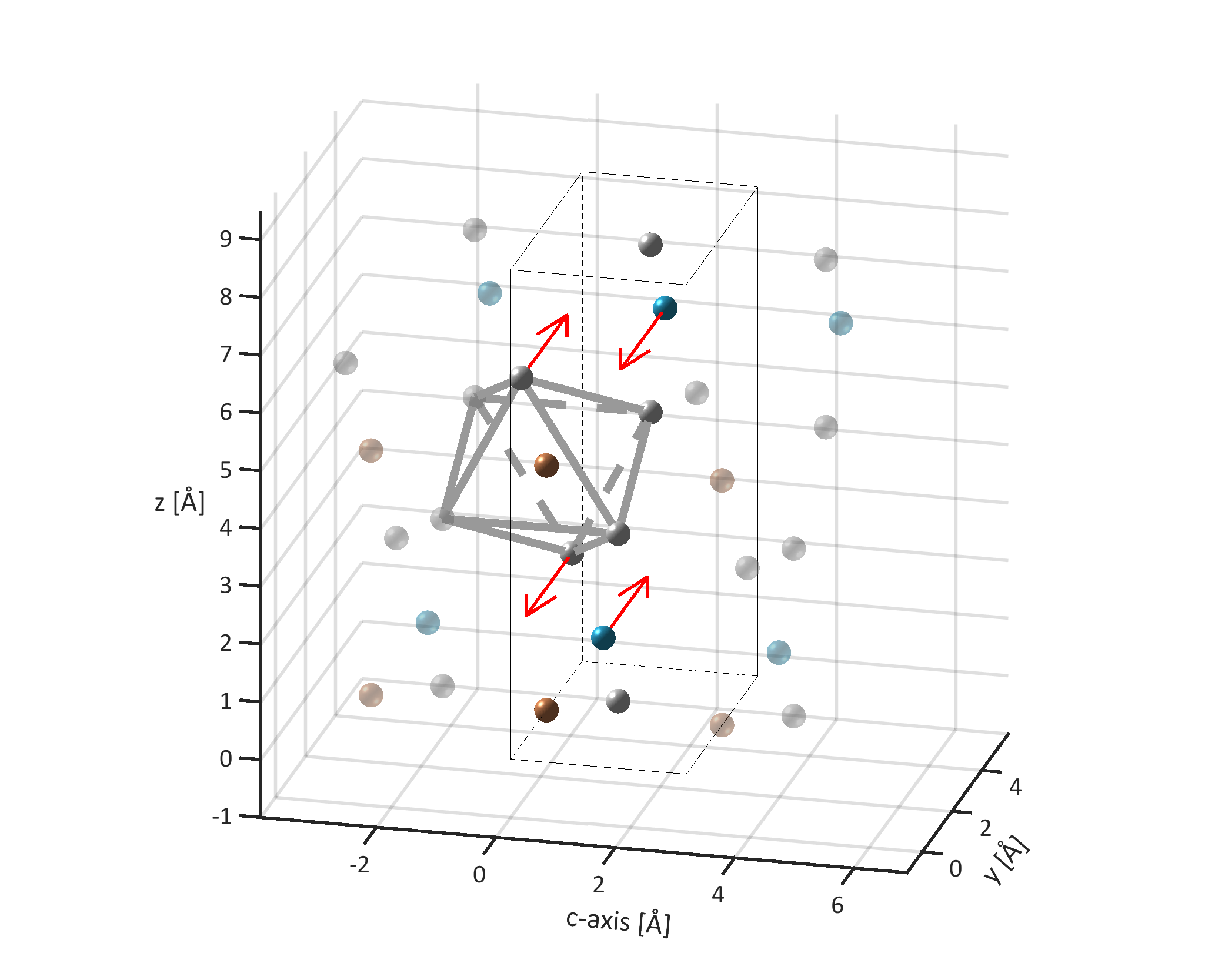}   
\caption{The sum of the force fields given in Table \ref{force_field_plus} and \ref{force_field_minus} is depicted by the red arrows which corresponds to the remaining force field on atoms when they have been displaced from equilibrium along the normal mode $B_{2u}$ by +$\Delta_x$ and -$\Delta_x$. The small spheres correspond to the equilibrium atomic positions One mesh is delimited by the thin gray lines and one octahedron is depicted by the thick gray lines.}
\label{force_field}
\end{figure}

\noindent After the structure relaxation one still finds two negative eigenvalues, however, this is consistent with the uncertainty of around $40$ cm$^{-1}$ that one can estimate by comparing the theoretical result with the fit of experimental data \cite{Popovic1995}.
We are mosly interested in the high frequency sector of the spectrum, in particular in the IR active mode $B_{2u}$ ($702$ cm$^{-1}$) that should be the mostly excited with a pump polarized along the c-axis (it is the highest frequency among the modes with the right symmetry).  
In order to estimate the anharmonic effects on the perturbation of the octahedron we displace the atoms along the normal mode $B_{2u}$ and compute the force field resulting on each atom. The results are presented in Tables \ref{force_field_plus} for a positive displacement along the c-axis (and in Table \ref{force_field_minus} for a negative displacement). The remaining force field, which is computed by the sum of the force fields for opposite displacements, is depicted in Figure \ref{force_field} and it shows that the apical oxygens are forced to move.

\begin{table}[h!]
\begin{center} 
\caption{Forces acting on atoms (cartesian axes, Ry/au) for a positive displacement (+$\Delta_x$) }
\label{force_field_plus}

\begin{tabular}{|c|c|c|c|c|}
\hline
 Atom &  Type & F$_x$ & F$_y$ & F$_z$\\ \hline
      1 &  O   &     0.01027863  &  -0.00012958  &  -0.00006228 \\ \hline
      2 &  O   &     0.01027863 &    0.00012958 &    0.00006228\\ \hline
      3 &  O   &     0.01027863  &  -0.00012789  &   0.00005886\\ \hline
      4 &  O   &     0.01027863  &   0.00012789  &  -0.00005886\\ \hline
      5 &  O   &    -0.08683023 &   -0.01712615 &    0.00000245\\ \hline
      6 &  O   &    -0.08683023  &   0.01712615 &   -0.00000245\\ \hline
      7 &  Cu  &    -0.01005315  &   0.00000000 &    0.00000000\\ \hline
      8 &  Cu  &    -0.01005329  &   0.00000000  &   0.00000000\\ \hline
      9 &  Ge  &     0.07632619  &   0.01659193  &   0.00000174\\ \hline
     10 &  Ge  &     0.07632619 &   -0.01659193  &  -0.00000174\\ \hline
\end{tabular}
\end{center}
\end{table}

\begin{table}[h!]
\begin{center} 
\caption{Forces acting on atoms (cartesian axes, Ry/au) for a negative displacement (-$\Delta_x$)}
  \label{force_field_minus}       
\begin{tabular}{|c|c|c|c|c|}
 \hline        
 Atom &  Type & F$_x$ & F$_y$ & F$_z$\\ \hline
      1 &  O   &    -0.01027890  &  -0.00012886 &   -0.00005982\\ \hline
      2 &  O   &    -0.01027890  &   0.00012886 &    0.00005982\\ \hline
      3 &  O   &    -0.01027885  &  -0.00012845 &    0.00006007\\ \hline
      4 &  O   &    -0.01027885  &   0.00012845 &   -0.00006007\\ \hline
      5 &  O   &     0.08683052  &  -0.01712782 &   -0.00000103\\ \hline
      6 &  O   &     0.08683052  &   0.01712782 &    0.00000103\\ \hline
      7 &  Cu  &     0.01005288  &   0.00000000 &    0.00000000\\ \hline
      8 &  Cu  &     0.01005286  &  0.00000000   &  0.00000000\\ \hline
      9 &  Ge  &    -0.07632564  &   0.01658938  &   0.00000389\\ \hline
     10 &  Ge  &    -0.07632564   & -0.01658938 &   -0.00000389\\ \hline

\end{tabular}
\end{center}
\end{table}

%%\bibliographystyle{apalike}
%%\bibliographystyle{abbrv}

%\bibliography{Biblio_vibration_control_dd_CuGeO3}

\end{document}